\documentclass[conference]{IEEEtran}
\makeatletter
\def\ps@headings{%
\def\@oddhead{\mbox{}\scriptsize\rightmark \hfil \thepage}%
\def\@evenhead{\scriptsize\thepage \hfil \leftmark\mbox{}}%
\def\@oddfoot{}%
\def\@evenfoot{}}
\makeatother
\usepackage{graphicx}
\usepackage{subfigure}
\usepackage{amsmath}
\usepackage{subfigure}
\usepackage{algorithm}
\usepackage{algorithmic}
\usepackage{bm}
\usepackage{ifpdf}

\begin{document}

\title{Efficient and Secure Key Extraction using CSI without Chasing down Errors}
\author{Jizhong Zhao$^{\dagger}$, Wei Xi$^{\dagger}$, Jinsong Han$^{\dagger}$,
Shaojie Tang$^{\ast}$, Xiangyang Li$^{\ast}$, Yunhao Liu$^{\S}$, Yihong Gong$^{\dagger}$, Zehua Zhou$^{\dagger}$\\
$^{\dagger}$School of Electronic and Information Engineering, Xi'an Jiaotong University, China \\
$^{\ast}$ Department of Computer Science, Illinois Institute of Technology, USA\\
$^{\S}$ School of Software and TNLIST, Tsinghua University, China\\
}

\maketitle

\begin{abstract}
Generating keys and keeping them secret is critical in secure
communications. Due to the ``open-air" nature, key distribution
is more susceptible to attacks in wireless communications.
An ingenious solution is to generate common secret keys by two
communicating parties separately without the need of key exchange
or distribution, and regenerate them on needs. Recently, it is
promising to extract keys by measuring the random variation in
wireless channels, \emph{e.g.}, RSS. In this paper, we propose an efficient Secret
Key Extraction protocol without Chasing down Errors, SKECE.
It establishes common cryptographic keys for two communicating parties in wireless
networks via the real-time measurement of Channel State Information
(CSI). It outperforms RSS-based approaches for key generation in
terms of multiple subcarriers measurement, perfect symmetry in channel,
rapid decorrelation with distance,
and high sensitivity towards environments.
In the SKECE design, we also propose effective mechanisms such as
the adaptive key stream generation, leakage-resilient consistence
validation, and weighted key recombination, to fully exploit the
excellent properties of CSI. We implement SKECE on off-the-shelf
802.11n devices and evaluate its performance via extensive experiments.
The results demonstrate that
SKECE achieves a more than $3\times$ throughput gain in the key generation from
one subcarrier in static scenarios,
and due to its high efficiency, a 50\% reduction on the communication
overhead compared to the state-of-the-art
RSS-based approaches.
\end{abstract}

\IEEEpeerreviewmaketitle

\section{Introduction}
Wireless networks are susceptible to various attacks due to the
``open air" nature of the wireless communication \cite{zhou1999securing}. Cryptographic key
establishment is a fundamental requirement for secure communication
to support confidentiality and authentication services. However,
it is difficult to ensure availability of a certificate authority
or a key management center in dynamic wireless environments \cite{chan2003random}.
It is necessary to have alternatives for key agreement between
wireless entities in a common channel \cite{lee2006distributed} \cite{liu2005establishing} \cite{amir2001exploring}.

One recent trend in this regard is to use physical-layer identification \cite{radio-telepathy}.
For example, received signal strength (RSS) becomes a
popular statistic of the radio channel and is used as the source of
secret information shared between two parties \cite{effkeyex}. The variation over
time of the RSS, caused by motion multipath fading,
can be quantized and used for generating secret keys.
Due to presence of noise and manufacturing variations,
the generated secret keys might be different, which are
corrected by information reconciliation. Finally, privacy amplification
is introduced to convert this bit-string into a uniformly
distributed string to make it secure enough.

However, RSS cannot work well in stationary scenarios due to
infrequent and small scale variations in the channel measurements.
To address this issue, we propose a secret key extraction based on the
inherent randomness of wireless channels.
In current widely used IEEE 802.11n networks, data is modulated
on multiple Orthogonal Frequency Division
Multiplexing (OFDM) subcarriers simultaneously.
Each network interface card (NIC) of the device can get a value of
Channel State Information (CSI) which describes the current condition
of the channel in each subcarrier \cite{crepaldicsi}.

Different from
RSS, CSI is a fine-grained value derived from the physical layer. It
consists of the attenuation and phase shift experienced by each
spatial stream on every subcarrier in the frequency domain.
In contrast to having only one RSS value per packet, NIC can obtain
multiple CSI values at one time. CSI provides other attractive
properties. First, it is very sensitive to location such
that two closely-placed receivers have very different readings
by the same sender. Second, its readings of a pair of sender
and receiver have a strong correlation. Third, it
presents an excellent quality of randomness. Due to these characteristics,
CSI is an ideal resource for secret key extraction.

In this paper, we present the design and implementation of CSI-based
Secret Key Extraction without Chasing down Errors. SKECE exploits channel
diversity to generate secret key using CSI. The contributions
of this work are summarized as follows.

1. We first give an insight into how CSI measurements improve the effectiveness and safety
of secret key extraction based on extensive real world measurements.
Our observation suggests that CSI possesses excellent symmetry in channels,
sensitivity to the environment, and rapidly decorrelates over a distance, which can work well in both
static and mobile scenarios, and effectively prevent the predictable channel attack.
Moreover, it can be measured from multiple subcarriers simultaneously, which significantly
improves the rate of key generation.

2. We propose an efficient and secure consistency validation method.
It avoids leaking any available information to attackers, when two parties
check the consistency of the generated bit streams. Additionally, it has a high precision
rate of consistency validation.

3. We creatively propose a weighted key recombination. This method can
efficiently recombine the mismatched bit streams into a consistent stream for two parties
without detecting which bits are mismatched. It reduces the communication overhead while
enhancing privacy and security by fully exploiting the special properties of OFDM modulation.

4. We evaluate SKECE through various experiments using off-the-shelf 802.11
devices in real indoor and outdoor scenarios.

The rest of this paper is organized as follows. Section 2
briefly reviews the related work. Section 3 presents real world
observations on 802.11n devices and the adversary
model.  The design of SKECE is elaborated in Section 4,
followed by performance evaluation in Section 5. We conclude the paper in Section 6.

\begin{figure*}[tbp]
\begin{minipage}[t]{0.49\linewidth}
\centering
\includegraphics[height=3.8cm]{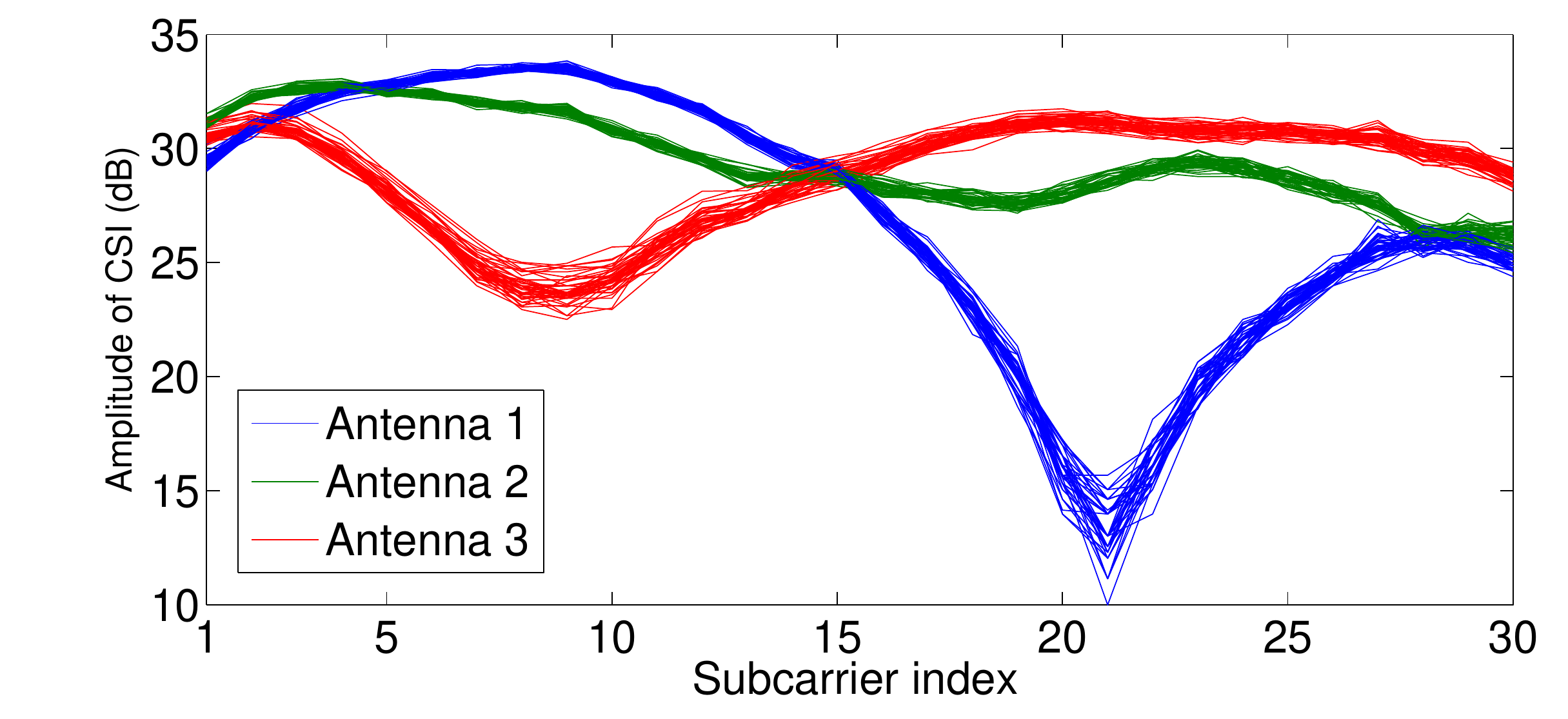}
\caption{Multiple readings for CSI measurements from 30 subcarriers}
\label{csi}
\end{minipage}
\hfill
\begin{minipage}[t]{0.49\linewidth}
\centering
\includegraphics[height=3.8cm]{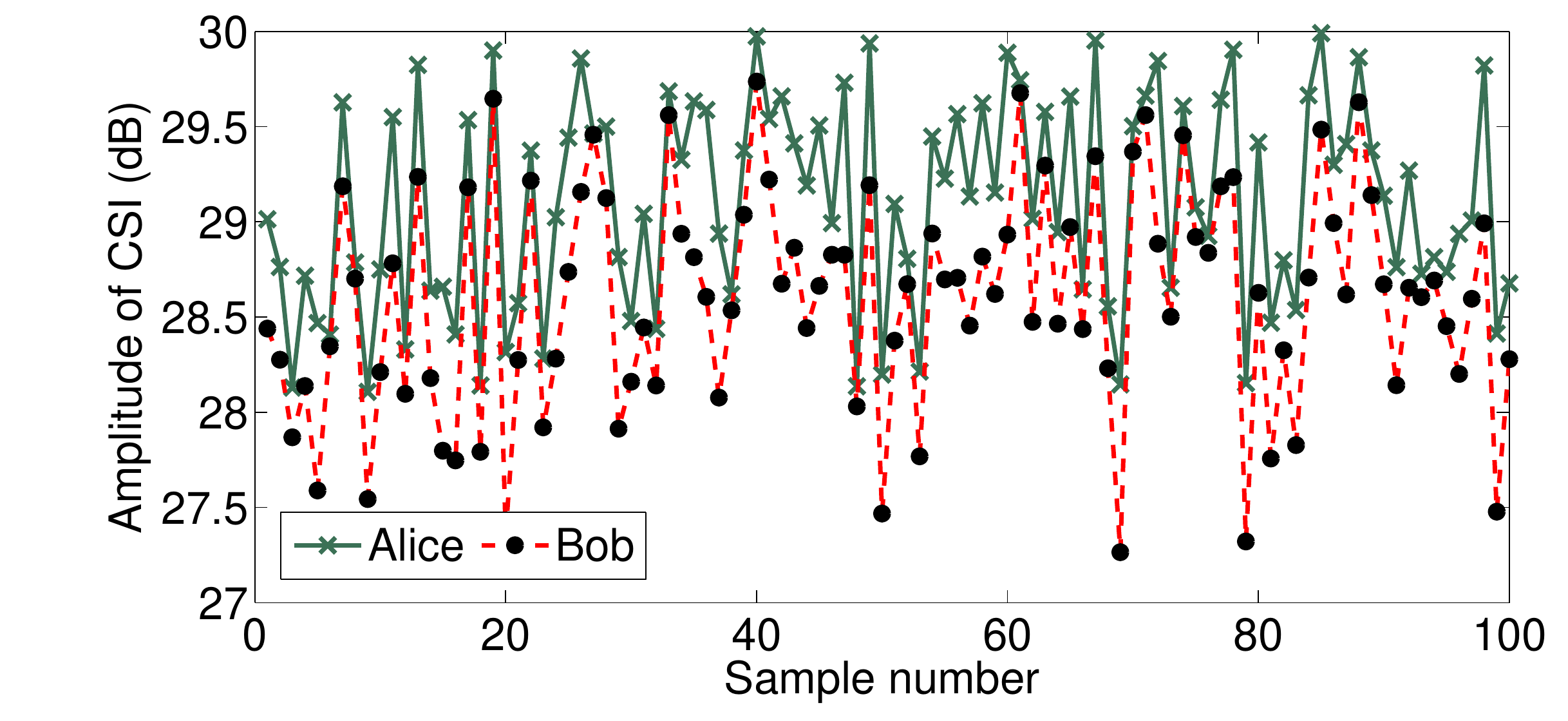}
\caption{CSI values in stationary environment}
\label{static}
\end{minipage}
\end{figure*}

\begin{figure*}[tbp]
\begin{minipage}[t]{0.49\linewidth}
\centering
\includegraphics[height=3.8cm]{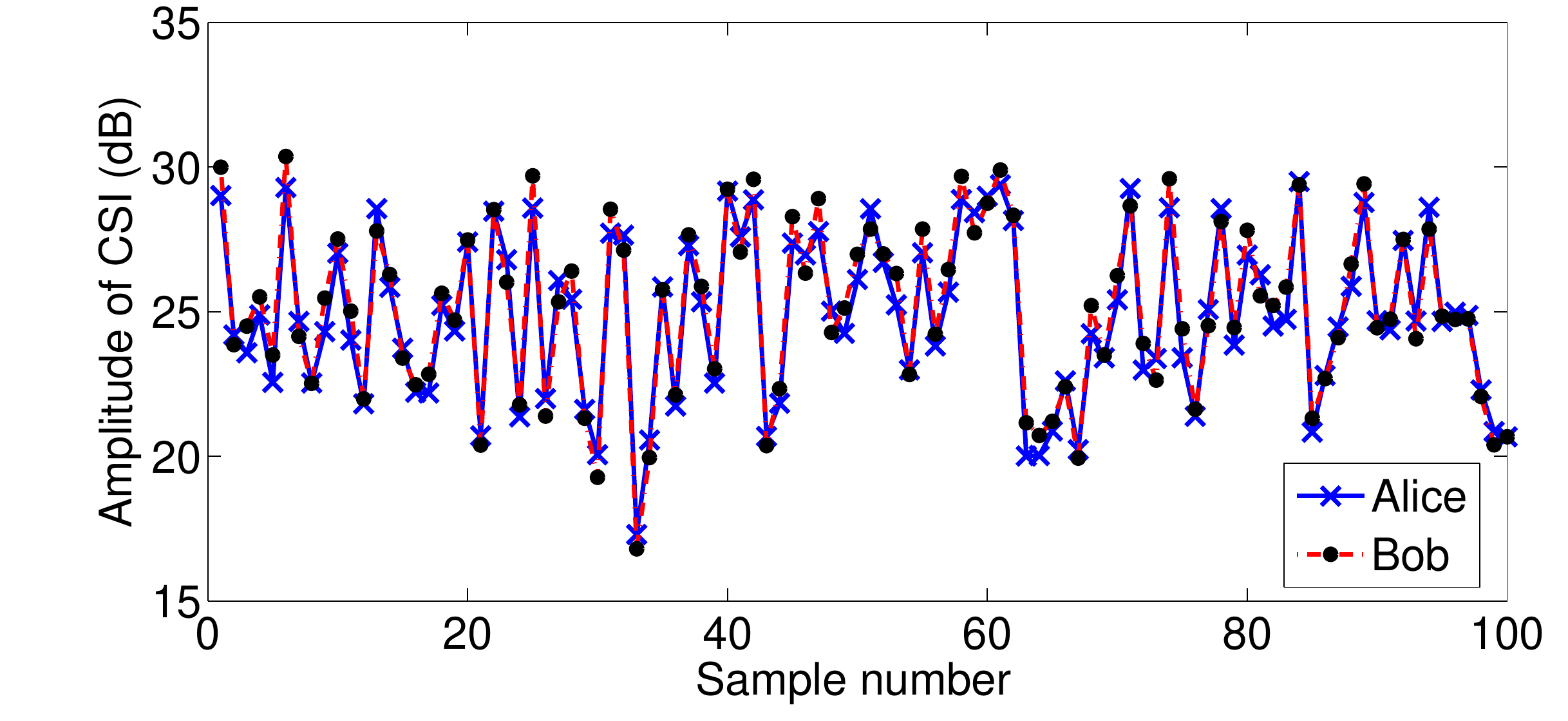}
\caption{Variations of CSI when Alice moves}
\label{move}
\end{minipage}
\hfill
\begin{minipage}[t]{0.49\linewidth}
\centering
\includegraphics[height=3.8cm]{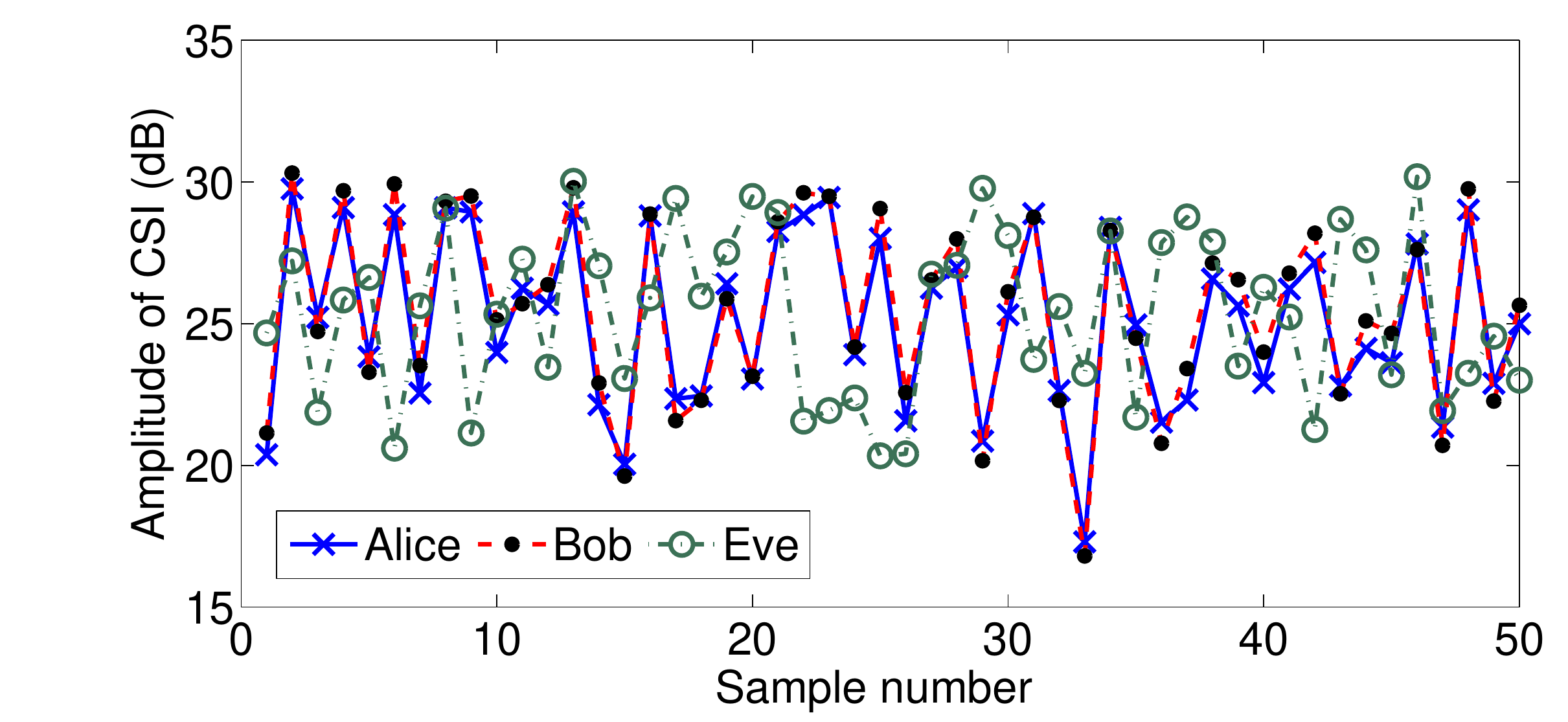}
\caption{Variations of CSI in different locations}
\label{eve}
\end{minipage}
\end{figure*}

\section{Related Work}

Encryption and authentication on communication between two parities in wireless networks is helpful for privacy and
sensitive data protection \cite{zhou1999securing} \cite{zhu2003establishing} \cite{zhu2004interleaved} \cite{li2010data}.
Extracting a shared secret key from the observation and processing of radio channel parameters has been
proposed to address this problem without resorting to a fixed infrastructure.

Typical secret key generation process consists of three phases: randomness exploration, information
reconciliation and privacy amplification \cite{ecd}. In the randomness exploration, quantization is used
to convert measurement values to information bits. A good quantizer can maximize the mutual
information between Alice and Bob without information leakage.  An algorithm is proposed in
\cite{effkeyex} to find such a quantizer.  The
information reconciliation process uses either error correcting codes \cite{fuzzy-ex}, or
interactive information reconciliation protocols, \emph{e.g.} Cascade \cite{Brassard1994}. The
universal hash functions are widely adopted in \cite{Maurer2003,wilhelm2009key,pr-one-way}  to enhance
privacy and security.

There are also many works on exploiting physical channel randomness feature to generate secret key
\cite{itsk,hershey1995unconventional,hassan1996cryptographic}. The authors in \cite{radio-telepathy} discuss the condition of generating secure keys and propose a solution to extract a secret key from unauthenticated wireless channels using  channel impulse response and amplitude measurements. The authors in \cite{effkeyex} summarize the processes needed for key extraction, give their choices of the methods in every process, and conduct extensive experiments to show the properties of
RSS in real environment.

It can also be exploited for device pairing
\cite{heartbeats2011proximate} and authentication
\cite{xiao2008using}.  Extracting secret keys over MIMO has
been introduced in \cite{wallace2010automatic}.

Previous works are mainly based on RSS, a coarse indicator of signal. As a fine-grained
indicator of channels, CSI draws increasing attentions
It can be measured using off-the-shelf 802.11n devices.
In this paper, we suggest that the
channel randomness can be further exploited through the channel diversity offered by CSI to
efficiently extract secret key.

\begin{figure}[htb]
\begin{minipage}[t]{1\linewidth}
\centering
\includegraphics[height=5cm]{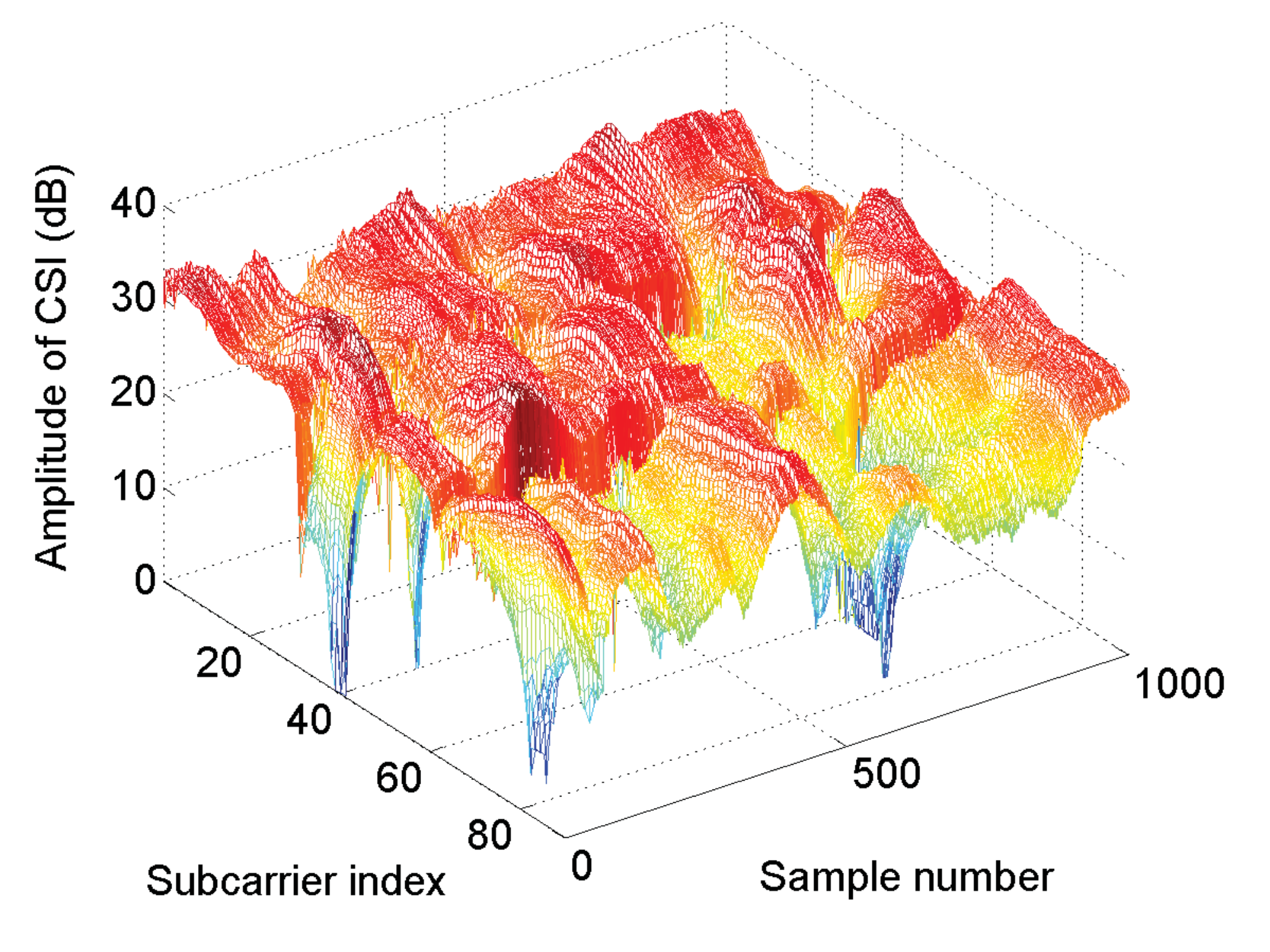}
\caption{Alice walks freely to measure CSI values from 90 subcarriers simultaneously using a network card with 3 antennas.}
\label{subca}
\end{minipage}
\end{figure}

\section{Preliminary Observation and Attack Models}
Orthogonal Frequency Division Multiplexing (OFDM) is a method
of Multicarrier Modulation. OFDM divides previous single-carrier
into a set of orthogonal sub-carriers, which can convert one
rapidly modulated wideband signal to many slowly modulated
narrowband signals.
Channel state information (CSI) describes the current channel¡¯s
condition, which can reveal the effect of scattering, fading and
power decay with distance. 802.11n protocol provides 30 pairs
of amplitude and phase information out of 56/114 sub-carriers. Each
pair of amplitude and phase describes the state information of a
sub-carrier. The available CSI can reflect the environment influences
on the signals transmitted from transmitter(s) to receiver(s):
\begin{equation}
    Y = HX + N
\end{equation}
\begin{displaymath}
    H=Y/X=he^{jw}
\end{displaymath}
where $X$ is transmit signal, $Y$ is received signal, $N$ is the noise, $H$
is the CSI: the channel response at the receiver in frequency domain,
$h$ and $w$ is the amplitude and phrase.

Utilizing off-the-shelf 802.11n wireless net card, CSI can be
collected. It reports the channel matrices for 30 subcarrier groups,
which is about one group for every 2 subcarriers at 20 MHz or one in 4
at 40 MHz. Although the driver does not directly provide functions
to get CSI, there are some open source tools on Linux platform that can be
used to collect CSI. We use Linux 802.11n CSI Tool and Intel 5300 wireless net card to spread out the
signal received by one antenna and to provide 30 pairs of amplitude and
phase CSI values to each antenna \cite{csitool}. Since 802.11n supports
MIMO, Fig. \ref{csi} shows the CSI measurements from three antennas of
one NIC.

In our experiments, we setup one laptop (named Alice) to
connect to the other laptop (named Bob). To establish a shared secret key, Alice and
Bob should measure the variation of the wireless channel at the same
time. However, typical commercial wireless transceivers are
half duplex, \emph{i.e.}, they cannot transmit and receive signals
simultaneously. We use \emph{ping} command to guarantee the time
between two directional channel measurements is small enough. The initiator
requests the receiver to immediately reply once receiving order.
The round-trip delay is always between 1ms-5ms.
Another laptop (named Eve) is introduced to overhear the packets
delivered  between Alice and Bob and measure the CSI variation.
Our observations for the feature of CSI is extracted into
following 4 subsections.

\subsection{Reciprocity of Radio Wave Propagation}
The multipath properties of a radio channel at any point
are identical on both directions of a link. Figure \ref{static}
and Figure \ref{move} shows the CSI values in static scenario
and mobile scenario. It is easy to see that Alice and Bob
have similar CSI variations.
\subsection{Temporal Variations in the Radio Channel}
Figure \ref{static} shows that CSI value is constantly changing
over time in static scenario. That is because the multipath
channel changes caused by any motion of people or objects in the
environment near the link.

\subsection{Spatial Variations}
The properties of a radio channel are unique to the locations
of the two endpoints of the link. Figure \ref{eve} shows that Eve
at a third location will measure a different CSI. This assertion
is supported by the well-known Jakes uniform scattering model,
which states that the received signal rapidly decorrelates over a
distance of roughly half a wavelength, and that spatial separation
of one to two wavelengths is sufficient for assuming independent
fading paths.

\subsection{Multiple Subcarriers of CSI}
IEEE 802. 11 a/g/n adopt OFDM to provide high throughput.
In OFDM, a channel is orthogonally divided into multiple
subcarriers. Figure \ref{subca} shows the multipath fading on a
mobile radio channel reflected in  90 subcarriers.
Measurements from multiple subcarriers can significantly
improve the key generation rate.

\subsection{Adversary Model}
For easy of exposition, we summarize our adversary model
as follows:

\begin{itemize}
  \item An adversary Eve can listen to all communications
between legitimate users.
  \item Eve can also measure channels between herself and other
parties anywhere she wants all the time.
  \item Eve is free to set intermediate objects between two parties
to affect their channels and derive some patterns known only to her.
  \item Eve knows the key extraction algorithm and parameters settings.
  \item Eve cannot prevent or modify any messages transmitted through the channel.
\end{itemize}

\section{Methodology}
Our CSI-based secret key extraction consists of three components:
adaptive bit streams generation, leakage-resilient consistency validation, and
weighted key recombination. For ease of presentation, Table \ref{variable} lists the
symbols and notations used in this paper.

\begin{table}
\centering
\caption{Variable Definitions}
\begin{tabular}{|c|c|l|} \hline
\textbf{Symbol}&\textbf{Definition}\\ \hline
$a$, $b$ & Alice, Bob \\ \hline
$m$ & the number of subcarriers ($m=30$ in our system) \\ \hline
$B_{ai}$ & a bit stream of Alice generated from her $i$-th subcarrier \\ \hline
$K_{ai}$ & a matched key of Alice generated from her $i$-th subcarrier \\ \hline
$K'_{ai}$ & a mismatched key of Alice generated from her $i$-th subcarrier\\ \hline
\end{tabular}
\label{variable}
\end{table}

\begin{figure}[t]
\centering
\includegraphics[width=9.5cm]{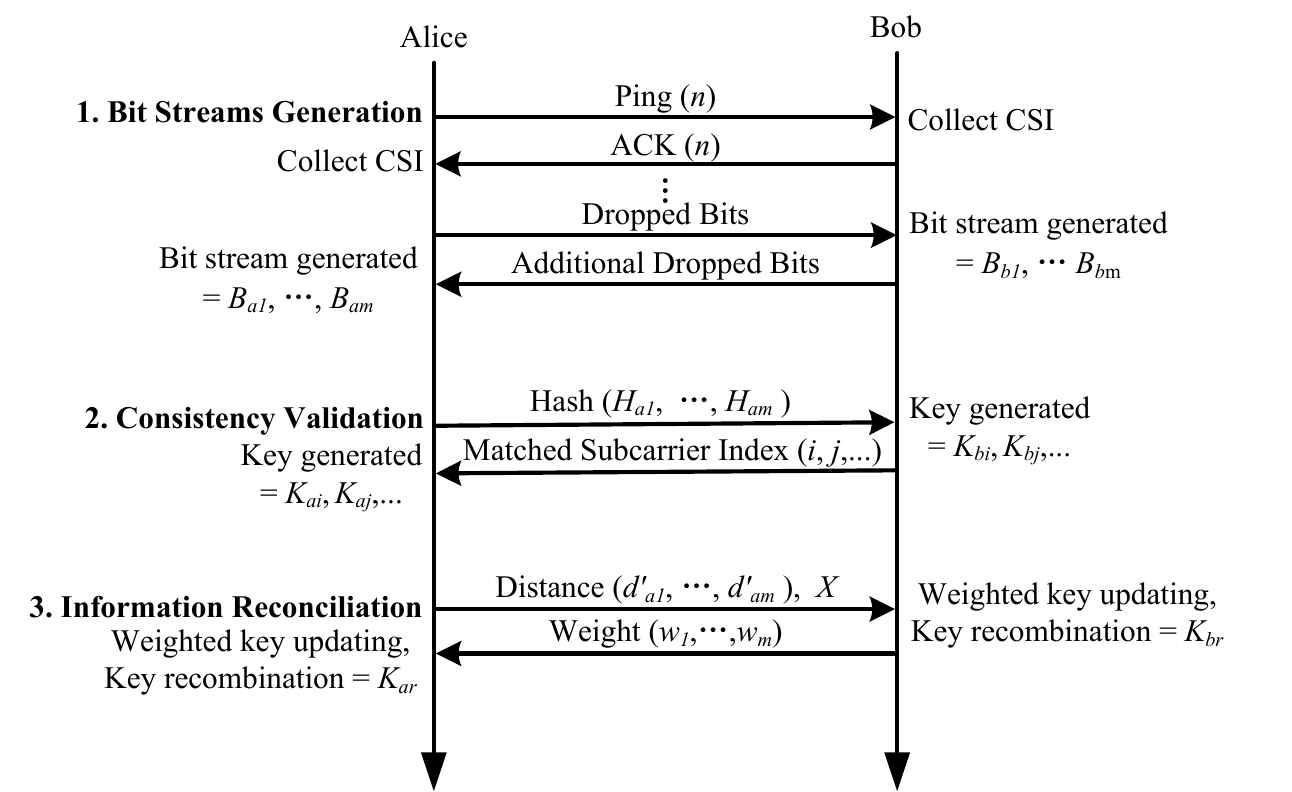}
\caption{Timing diagram for SKECE}
\label{timing}
\end{figure}

\subsection{Adaptive Bit Streams Generation}

To establish a shared secret key, Alice and Bob measure the variations
of the wireless channel between them over time by sending
probes to each other and measure the CSI values.
Ideally, Alice and Bob should both measure the CSI values at the
same time. However, typical commercial wireless transceivers are
half duplex, \emph{i.e.}, they cannot transmit and receive signals
simultaneously. Thus, Alice and Bob can only measure the radio channel
in one direction at a time. As long as the time between
two directional channel measurements is much smaller than the change rate
of channel, they will have similar CSI variations.

\subsubsection{Converting the channel measurements to bits}
Alice and Bob must convert their respective sequences of
channel estimation (\emph{i.e.}, the amplitude of CSI $S_a(t_1,...,t_n)$
and $S_b(t_1,...,t_n)$) into
identical bit-strings (0-1 sequences $B_a$ and $B_b$)
to be used as cryptographic keys. The bit-string should be
(1) \emph{Sufficient long}, ranging from 128 bits to 512 bits
being the length of keys commonly used in symmetric cipher
and (2) \emph{Statistically random}, resilient to statistical
defeats that could be exploited by attackers.

Many quantization methods have been proposed in previous
works. For example work in [2] partitions the CSI measurements to multiple
parts and performs quantization in every parts.

The quantizer we used is described as
follows. (i) Alice and Bob calculate two adaptive thresholds $q_+$
and $q_-$ independently such that $q_+ =\mu_{S(t_1,...,t_n)}+\alpha*\sigma_{S(t_1,...,t_n)}$
and $q_- = \mu_{S(t_1,...,t_n)} - \alpha * \sigma_{S(t_1,...,t_n)}$, where $\mu$
and $\sigma$ are the mean  value and standard deviation of $S(t_1,...,t_n)$, $\alpha \geq 0$.
(ii) Alice and Bob parse their CSI measurements and drop CSI estimates
that lie between $q_+$ and $q_-$ and maintain a list of indices to track
the CSI estimates that are dropped. They exchange their list of dropped
CSI estimates and only keep the ones that they both decide not to
drop. (iii) Alice and Bob generate their bit streams by extracting
a ``1" or a ``0" for each CSI estimate if the estimate lies above $q_+$ or
below $q_-$, respectively.

\subsection{Leakage-resilient Consistency Validation}
Alice and Bob extract the bit streams from the CSI
measurements they collect using quantizers. There might be
some differences in the corresponding bits between two streams.
They arise due to three factors: presence of noise and
interference, manufacturing variations, and the half-duplex
mode of communication between transceivers.

It is crucial for Alice and Bob to validate the consistency of
their bit streams without exposing available information
to the channel that can be overheard by Eve.

One-way function is one that is easy to
compute on every input, but hard to invert given the image of a random input.
One-way function is ideal for Alice and Bob checking
the consistency of keys without revealing information to malicious users.
We use SHA-1 to hash the bit streams, which is highly secure and most
widely used of one-way functions.

Alice sends the SHA-1 hash results ($H_{a1}, ..., H_{am}$) of her bit streams
to Bob, where $m$ is the number of subcarriers. Bob hashes
his bit streams in the same way, and compares the result with Alice's value.
The same hash results indicate the same original bit streams and vice versa.
The same bit stream is thereby used as the secret-key, denoted as $K_j$.
On the other hand, the different bit streams need to be corrected,
denoted as $K'_{ai}$ and $K'_{bi}$, respectively.

However, SHA-1 produces a 160-bit message digest. Directly transmitting
it may incur high communication cost. Fortunately, SHA-1 has the following
feature: even a small change in the message will, with
the overwhelming probability, result in a completely different hash. Due to
the avalanche effect, it can be considered that for two
different bit streams even if only one bit is different,
each bit in the SHA-1 hash values will have 50\%-50\% chance of being different.

Due to this feature, it is unnecessary to verify entire hash values
of two bit streams. Checking a small portion of hash values may have a
high probability to find the difference between two different bit streams.
The relationship between checking length $r$ and the correct probability $\gamma$ can
be described as follows:

\begin{equation}
1-(\dfrac1 2)^r\geq \gamma
\end{equation}
\begin{displaymath}
r=\lceil \log_{1/2}{(1-\gamma)}\rceil
\end{displaymath}

In our system, we set $r=6$ to achieve above 98\% correctness checking of detection results.

\subsection{Weighted Key Recombination}
If the bit streams extracted from all $m$ subcarriers are mismatched,
they cannot be used as the secret key. In this case, we perform reconciliation
to extract a consistent bit stream for Alice and Bob based on $m$ extracted
bit streams.Traditional information reconciliation techniques (\emph{e.g.} the error correcting
code and interactive information reconciliation) should first exclude matched
bits to shrink the parts of the streams containing mismatched bits using parity
checking or Hamming distance. The two parties permute their bit streams randomly,
divide the stream into blocks, and check the parity of each block. If the parity is
different between two sides, they repeat above procedure using a binary search
until the size of blocks is so small that attempting to replace a few bits may correct
the mismatch bits.

In contrast, using CSI can generate m bit streams from m subcarriers and most bits
of each stream are consistent as we have seen in Section III. We thereby propose a
weighted key recombination method. The key idea is very simple. We just let Alice and
Bob randomly pick up bits from the $m$ bit streams and recombine them into two new
bit streams $B_{ar}$ and $B_{br}$ without exchanging any information. Alice and Bob
random select the bits in same positions from the corresponding bit stream. The newly
generated bit steams have high probability to be matched.
Alice and Bob check the bit stream via consistency validation. If $B_{ar}$ is unequal  to $B_{br}$,
they will recombine the key again. The following two components realize our method.

\subsubsection{Difference degree detection}
Difference degree detection estimate how different the two bit streams $K'_{ai}$ and $K'_{bi}$ are.
Alice generates a random number $X$, and obtains the editor distance $d_{ai}$
between $K'_{ai}$ and $X$. Then she sends ($d_{a1}$,...,$d_{am}$) and $X$
to Bob. Bob compares its distance $d_{bi}$ with $d_{ai}$. The difference $\breve{d_i}$ between $d_{ai}$ and $d_{bi}$ reflects the difference between $K'_{ai}$ and $K'_{bi}$.
A larger $\breve{d_i}$  implies more difference between $K'_{ai}$
and $K'_{bi}$.

Since Alice and Bob always generate very similar bit streams,
$\breve{d_i}$ is quite small in most situations.
The value modulo $d_{ai}$ and $d_{bi}$ are enough to detect the difference of editor
distances of two parties.
We use $d'_{ai}$ instead of $d_{ai}$ to compute the distance
as follows:

\begin{equation}
d'_{ai}=d_{ai} \ \ mod \ \ \theta
\end{equation}
\begin{equation}
\tilde{d}'_i= |d'_{ai}-d'_{bi}|
\end{equation}

Choosing appropriate $\theta$ can reduce the communication cost and
avoid leaking too much information to the adversary Eve.
In our experiment, we set $\theta=5$.

\subsubsection{Secret key recombination}
Assume we randomly pick $l_i$ bits from $K_{ai}'$ and $K_{bi}'$ to generate
the new bit stream, the probability that all those $l_i$ bits are matched in $K_{bi}'$ is

\begin{equation}
\Pr_i(l_i) = (1-\frac{\hat{d_i}}{L})(1-\frac{\hat{d_i}}{L-1}) \cdots (1-\frac{\hat{d_i}}{L-l_i})
\end{equation}
where $L$ is the length of the key, $\hat{d_i}$ is the actual number of mismatched bits between $K_{ai}'$ and $K_{bi}'$.
Thus the overall probability to successfully generate a matched bit stream of length $L$ within $k$ rounds is

\[1- {\left(1-\prod_{i=1}^m \Pr_i(l_i)\right)}^k\]

We introduce a weight $ \omega_i$ to decide how
many bits should be selected from each one of $m$ bit streams.
The weight of each bit stream is computed as follows:
\begin{equation}
\omega_i =\dfrac{\theta-\tilde{d}'_i}{\sum_{j=1}^m (\theta-\tilde{d}'_j)}
\end{equation}

Then, the number $l_i$ of bits picked up from bit stream $B_i$ used to recombine
a new bit stream is
\begin{equation}
l_i=\lceil L*\omega_i\rceil.
\label{eq6}
\end{equation}

Indeed, $\tilde{d}'_i$ is proportional to $\hat{d_i}$, the value of $\omega_i$ implies a
match quality of two corresponding
bit stream $B_{ai}$ and $B_{bi}$. Following Equation \ref{eq6}, SKECE picks
more bits from those streams with more matched bits between Alice and Bob.

\section{Evaluation}

We conduct our experiments in a wide variety of environmental
settings and under different scenarios. The configuration is described
in Section 3. In our experiments, there are six scenarios as shown in Table \ref{scenario}.

\textbf{Scenario A:} Alice, Bob, and Eve put their laptops
on a boardroom table in a meeting room. Alice and Eve are 1.5m apart.
It's a quiet room without subject moving.

\textbf{Scenario B:} Alice, Bob, and Eve stay in an office. Alice and
Eve are 3m apart.There are various electronic devices and some walking people
in the room.

\textbf{Scenario C:} Alice, Bob, and Eve are in a meeting room. Alice and
Eve remain still separated with a distance of 3m. Bob walks
freely in the room.

\textbf{Scenario D:} It is similar to scenario \emph{C}, except that Alice
and Eve are 10cm apart from each other.

\textbf{Scenario E:} Alice, Bob and Eve stay in a garden. Alice and Eve
put their laptops on one bench separated with a distance of 10cm.
Bob's laptop is put on the other bench.

\textbf{Scenario F:} Alice, Bob and Eve walk freely on crowded street. Alice
and Eve keep a 3m distance from each other.

\begin{figure}[t]
\begin{minipage}[t]{0.49\linewidth}
\centering
\includegraphics[height=3.5cm]{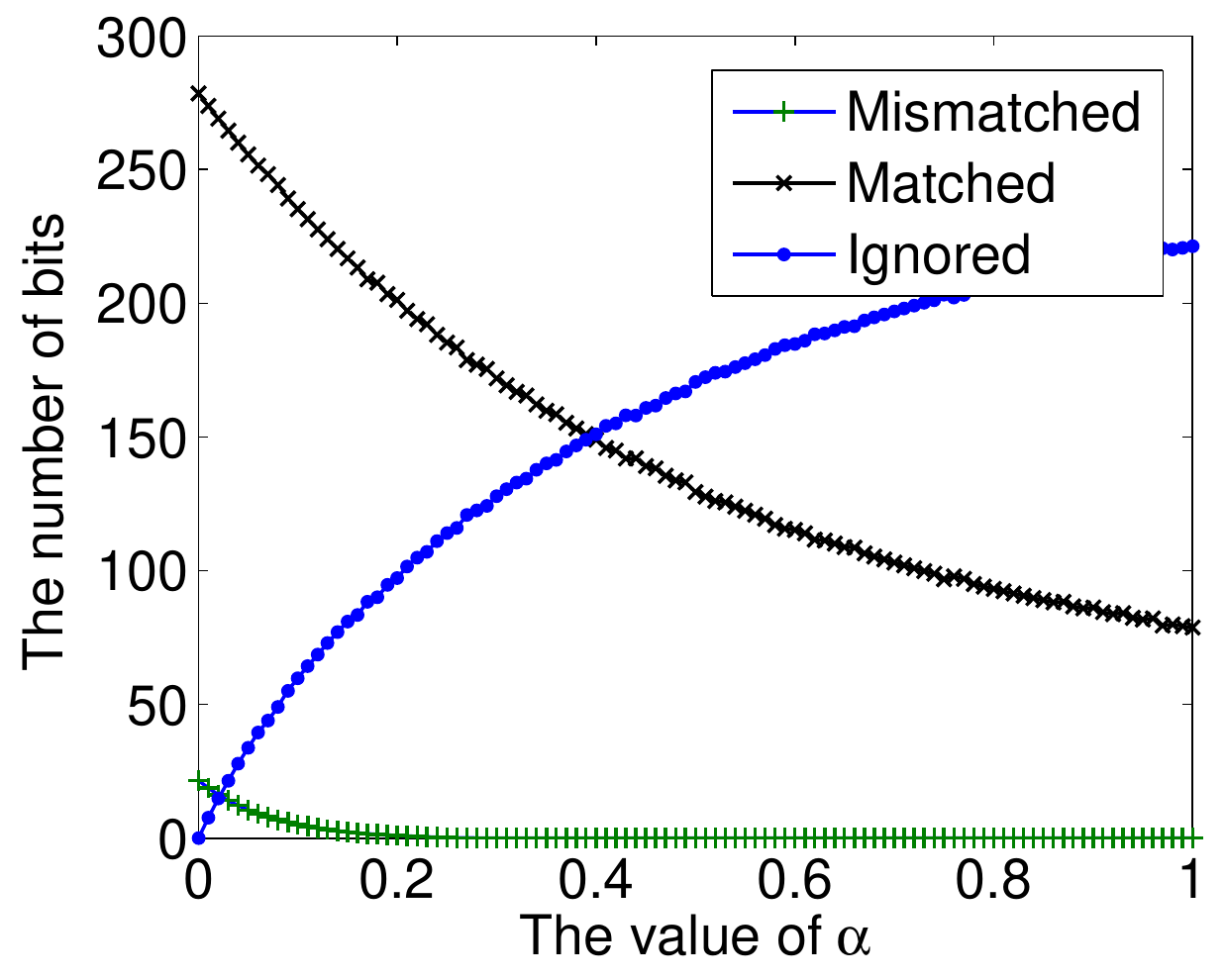}
\caption{Impact of $\alpha$}
\label{alpha}
\end{minipage}
\hfill
\begin{minipage}[t]{0.49\linewidth}
\centering
\includegraphics[height=3.5cm]{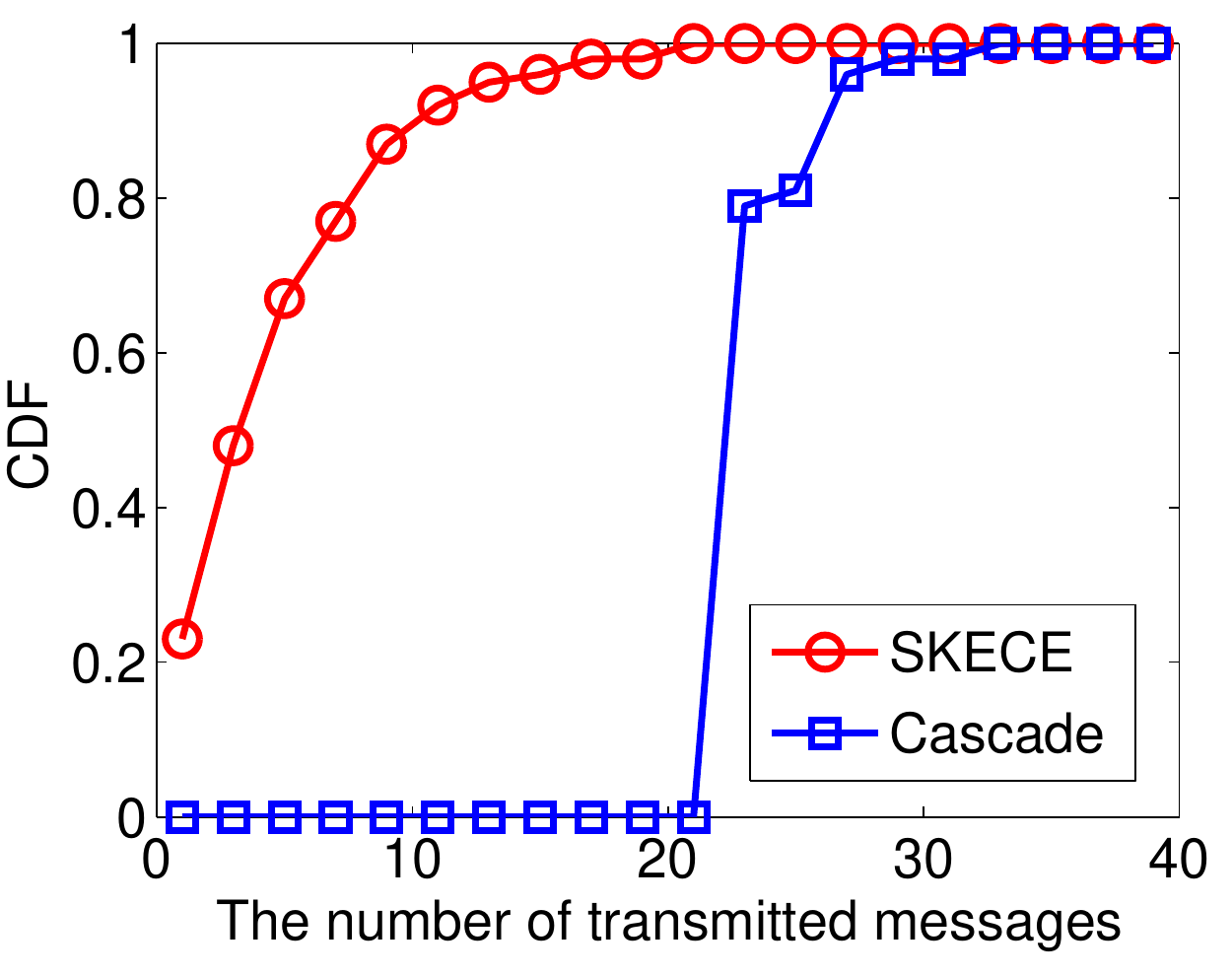}
\caption{Communication overhead}
\label{cdf}
\end{minipage}
\end{figure}

\begin{table}\centering
\begin{tabular}{|c|c|c|c|c|}
\hline
Index & Status & $Distance_{ae}$  & Environment       \\ \hline
A     & Static & 1.5 m     & Indoor,           \\ \hline
B     & Static & 3 m       & Indoor, Complex   \\ \hline
C     & Mobile & 3 m       & Indoor            \\ \hline
D     & Mobile & 10 cm     & Indoor            \\ \hline
E     & Static & 10 cm     & Outdoor           \\ \hline
F     & Mobile & 3 m       & Outdoor, Complex  \\ \hline
\end{tabular}
\caption{Parameters of datasets}
\label{scenario}
\end{table}

\begin{figure*}[tbp]
\subfigure[CSI measurements]{
\begin{minipage}[t]{0.49\linewidth}
\centering
\includegraphics[height=3.8cm]{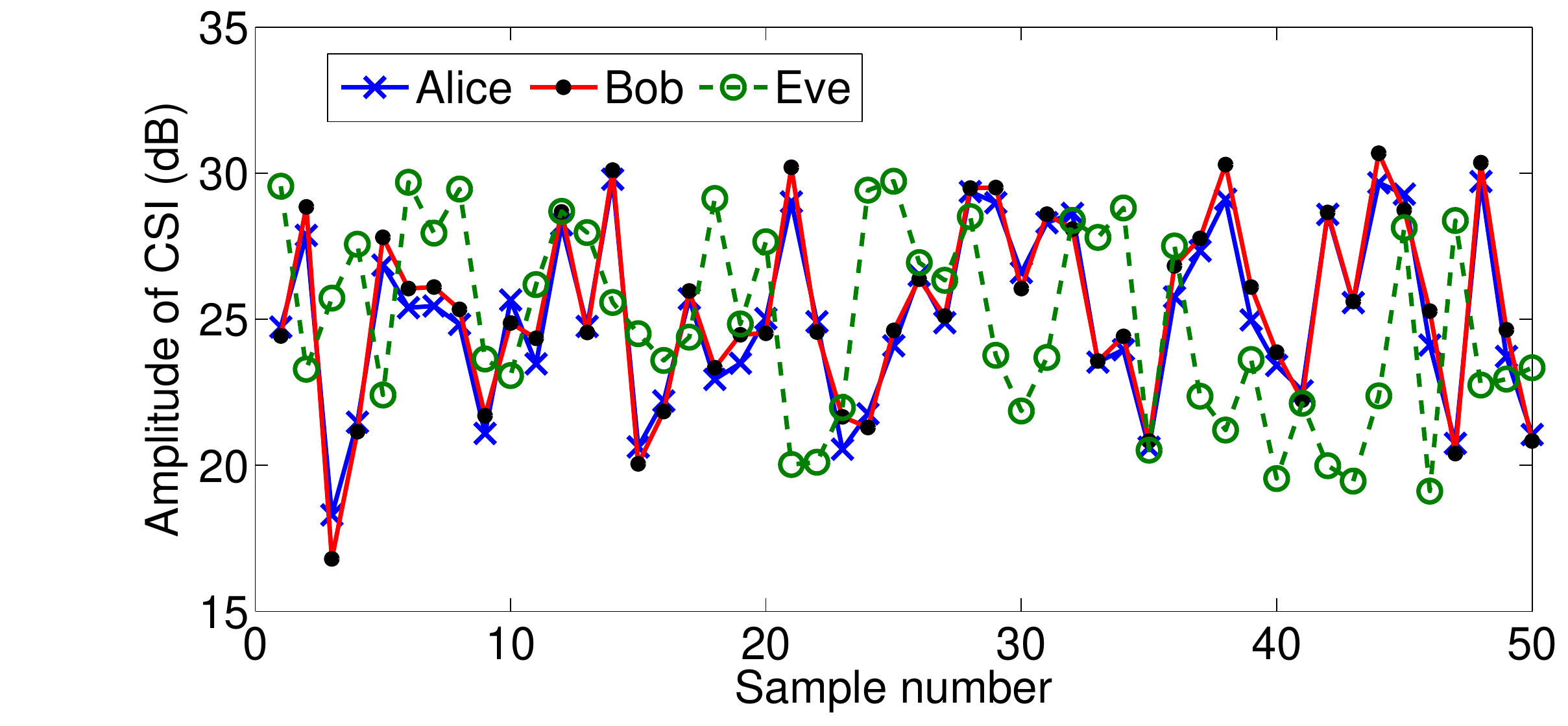}
\end{minipage}}
\hfill
\subfigure[RSS measurements]{
\begin{minipage}[t]{0.49\linewidth}
\centering
\includegraphics[height=3.8cm]{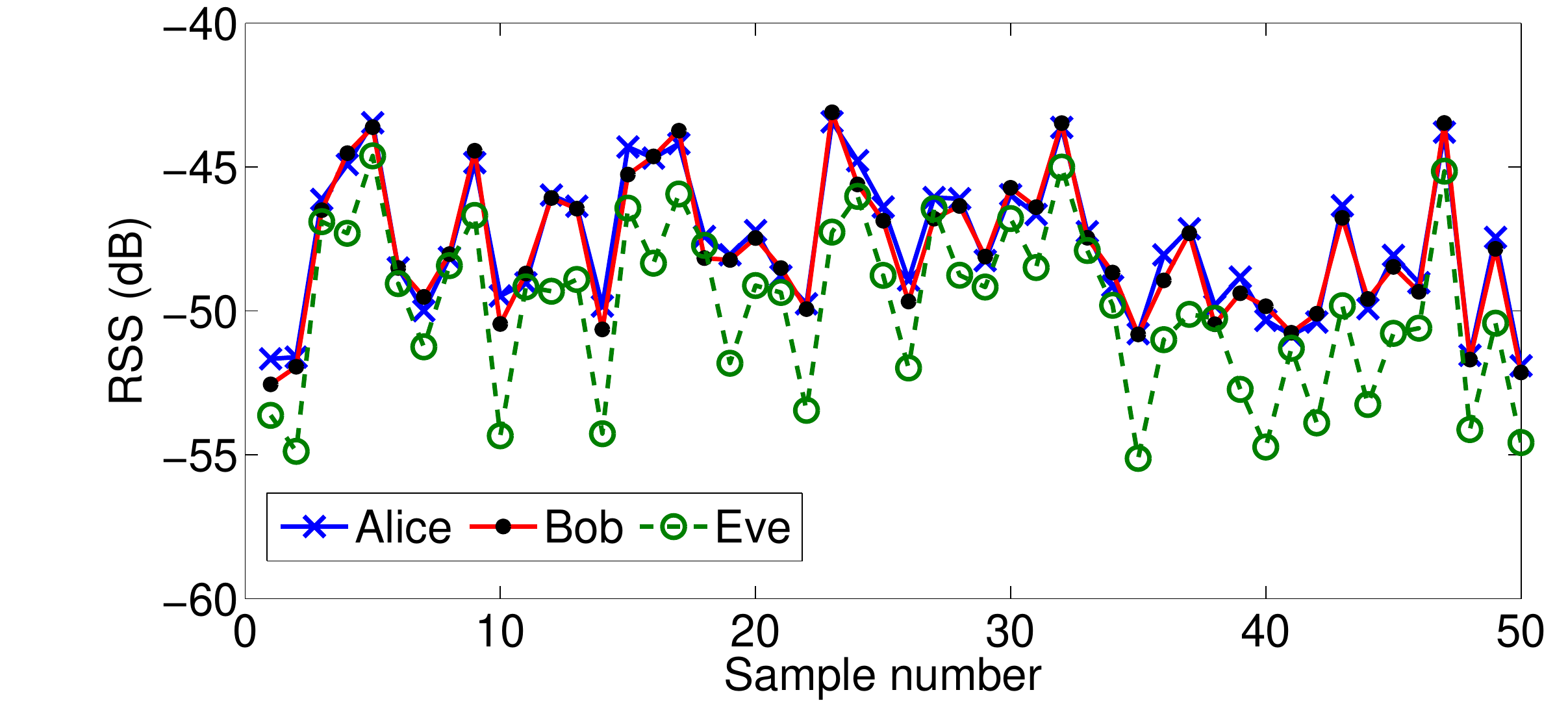}
\end{minipage}}
\caption{Measurements while walking in a meeting room}
\label{mobile_D}
\end{figure*}

\subsection{Converting the Channel to Bits}
As mentioned before, $\alpha$ affects the performance of
bit streams extraction from CSI measurements. Figure \ref{alpha}
shows the impact of $\alpha$ in scenario \emph{C}.
For ease of description, we divide the bits into three kinds:
ignored, mismatched and matched. An ignored bit is the dropped
bit whose value ranges from $q_-$ to $q_+$. A mismatched bit is the one
extracted as ``1'' in one party and extracted as ``0''
in the other party. A matched bit is the one that two parties
agree upon.

We investigate the variation of three kinds of bits extracted
from 300 probes by increasing $\alpha$ from 0 to 1.
Obviously, a smaller $\alpha$ improves
the rate of bits generation but increases the mismatched
bits ratio as well. With a larger $\alpha$, fewer bits can
be extracted, while the mismatched bits ratio reduces.

Specifically, the bit error should be carefully deal with.
If the sequence $B_{ai}$
is different from $B_{bi}$ even by a single bit, then the
two bit streams cannot be used as cryptographic keys.
As a result, two parties must reconciliate with each other.
High probability bit error will increase the
difficulty on information reconciliation. Therefore, we prefer
to choose a larger $\alpha$ to keep low mismatched bits ratio.

However, a too large $\alpha$ will seriously influence the rate
of bits generation. It is a fundamental trade-off in the selection
$\alpha$ that affects the rate and the probability of error in opposing
ways. An appreciate $\alpha$ can reduce the mismatched bits ratio
at acceptable of key generation. From the figure we can see that $\alpha$ greater
than 0.4 can reduce the bit errors to 0. In our experiment, we
set $\alpha=0.4$ in mobile scenarios and $\alpha=0.7$ in static
scenarios.

\subsection{Communication Overhead}
The communication overhead is a major concern from both the performance
and security perspectives, because the eavesdropper can overhear all communications
among legitimate users. We measure the number of messages transmitted between
Alice and Bob in each reconciliation process for SKECE and a typical RSS based approach,
Cascade used in \cite{Brassard1994, effkeyex}.

In the simulation, both Alice and Bob have 30 different bit streams, and each stream
has 300 random bits for SKECE. For Cascade, both the parties have one bit stream with
300 random bits. We randomly set 1 to 3 bits mismatched between two sides on their
corresponding streams. We then measure the number of transmitted messages for
achieving consistent 300 bits. Note that the number of bits for reporting differences in the
reconciliation is very small, \emph{e.g.} 1 bit used for parity checking plus several noise-padding bits
or 6 bits for SKECE. The length of data field in the message packet is much shorter than other fields,
\emph{e.g.} preamble 32 bytes, address $6\times3$ bytes. Thus, the dominated communication
overhead of reconciliation processes lies on the amount of delivered messages.

Figure \ref{cdf}  reports the comparison between two approaches.
Clearly, SKECE outperforms Cascade. For SKECE, more than 80\% cases need no more than 10
messages to achieve the consistence, while Cascade needs to deliver at least 20 messages
in most rounds. This is because that SKECE exploits the excellent properties of CSI to utilize
the matched bits to recombine the bit stream instead of adopting time-consuming detection
and correction mechanisms in the reconciliation process. \textbf{The result indicates that SKECE can
reduce 50\% communication overhead and benefit the efficiency improvement a lot.}

\subsection{Mobile Endpoints}
Considering that the mobility is an inherent property of wireless networks,
we evaluate the performance of key extraction in scenario \emph{D}.
Figure \ref{mobile_D} compares the channel estimation of CSI with that of RSS
in a meeting room.

The channel often varies with a wide variation window both in CSI (17dB
to 30dB) and RSS (-55dB to -42dB). Alice and Bob have a high degree
of reciprocity. This experiment shows that the mobility in indoor
setting helps achieve fast secret key extraction from the channel
measurements, which increases the inherent entropy of the measurements and
improves the reciprocity of the channel.

It is also interesting to see that Eve's observation is quite different from
Alice's and Bob's in CSI while she obtains some similarities with Alice and
Bob in RSS. That is because CSI has more decorrelation over a distance.

\begin{figure*}[tbp]
\subfigure[CSI measurements]{
\begin{minipage}[t]{0.49\linewidth}
\centering
\includegraphics[height=3.8cm]{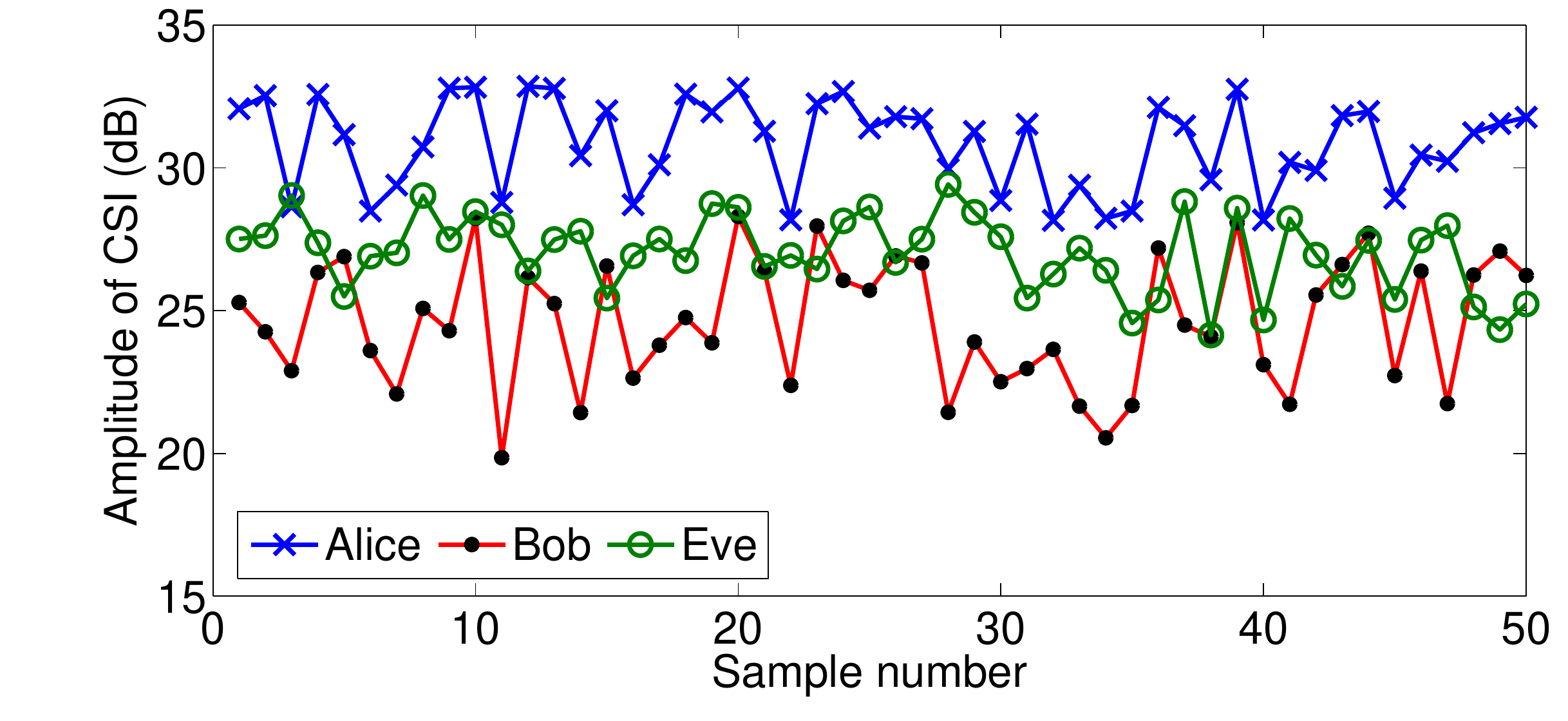}
\end{minipage}}
\hfill
\subfigure[RSS measurements]{
\begin{minipage}[t]{0.49\linewidth}
\centering
\includegraphics[height=3.8cm]{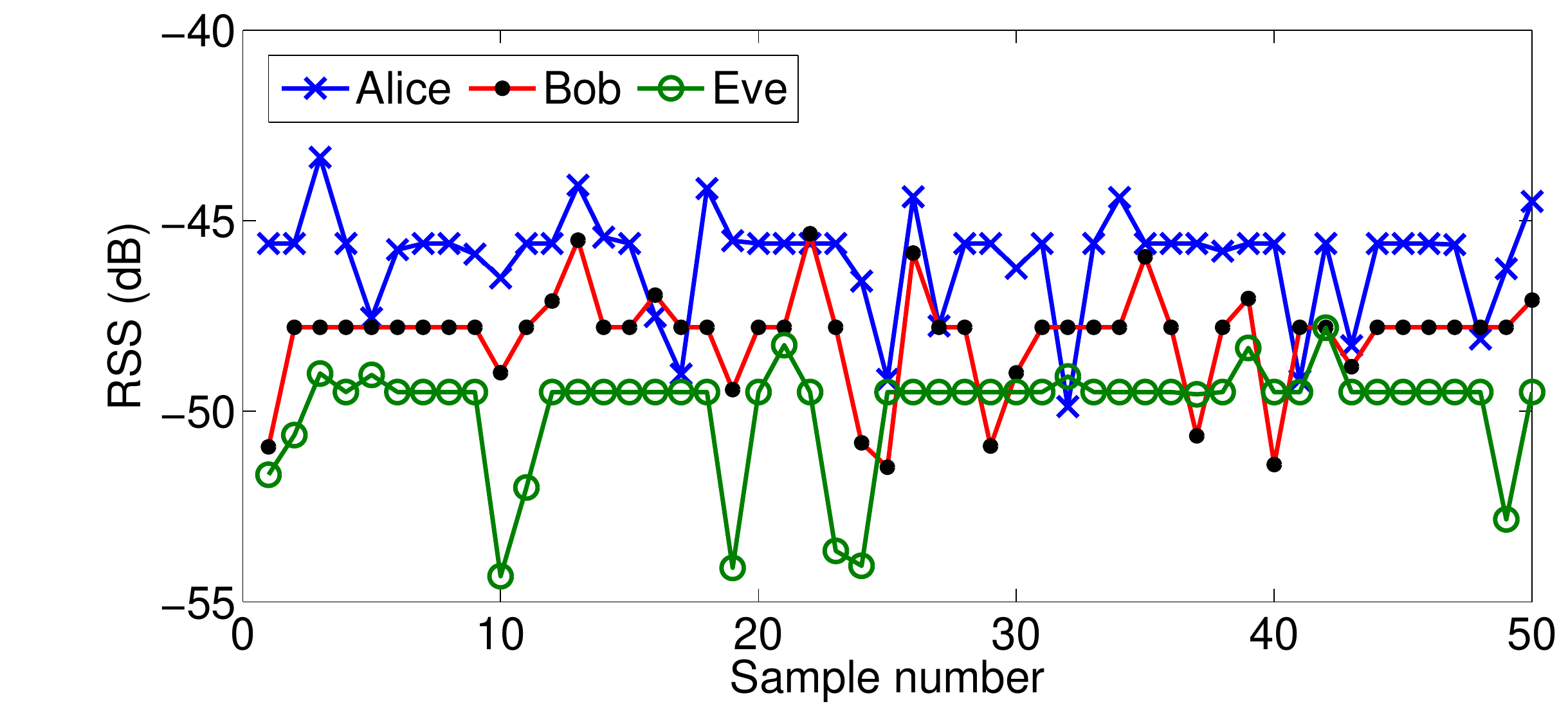}
\end{minipage}}
\caption{Measurements in an office}
\label{office}
\end{figure*}

\begin{figure*}[tbp]
\subfigure[CSI measurements]{
\begin{minipage}[t]{0.49\linewidth}
\centering
\includegraphics[height=3.8cm]{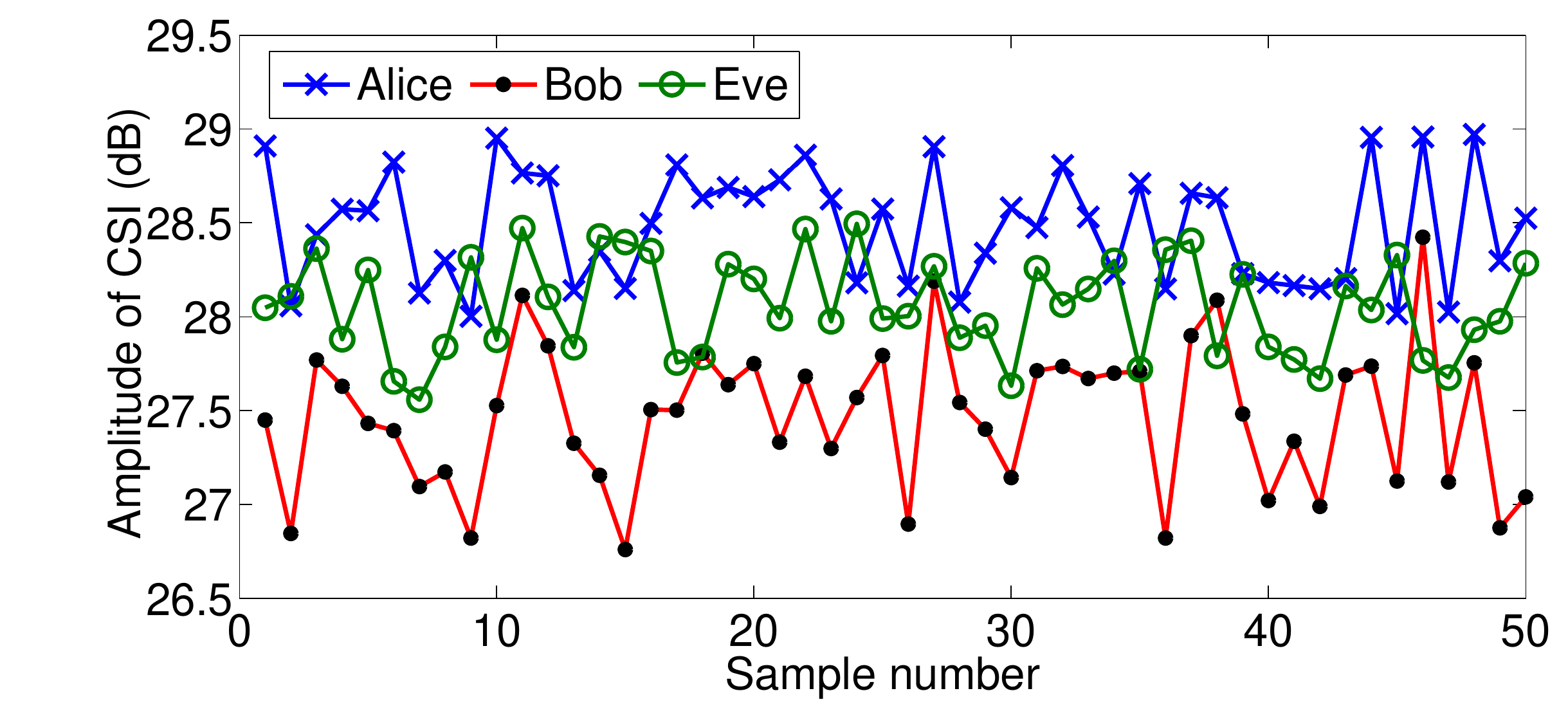}
\end{minipage}}
\hfill
\subfigure[RSS measurements]{
\begin{minipage}[t]{0.49\linewidth}
\centering
\includegraphics[height=3.8cm]{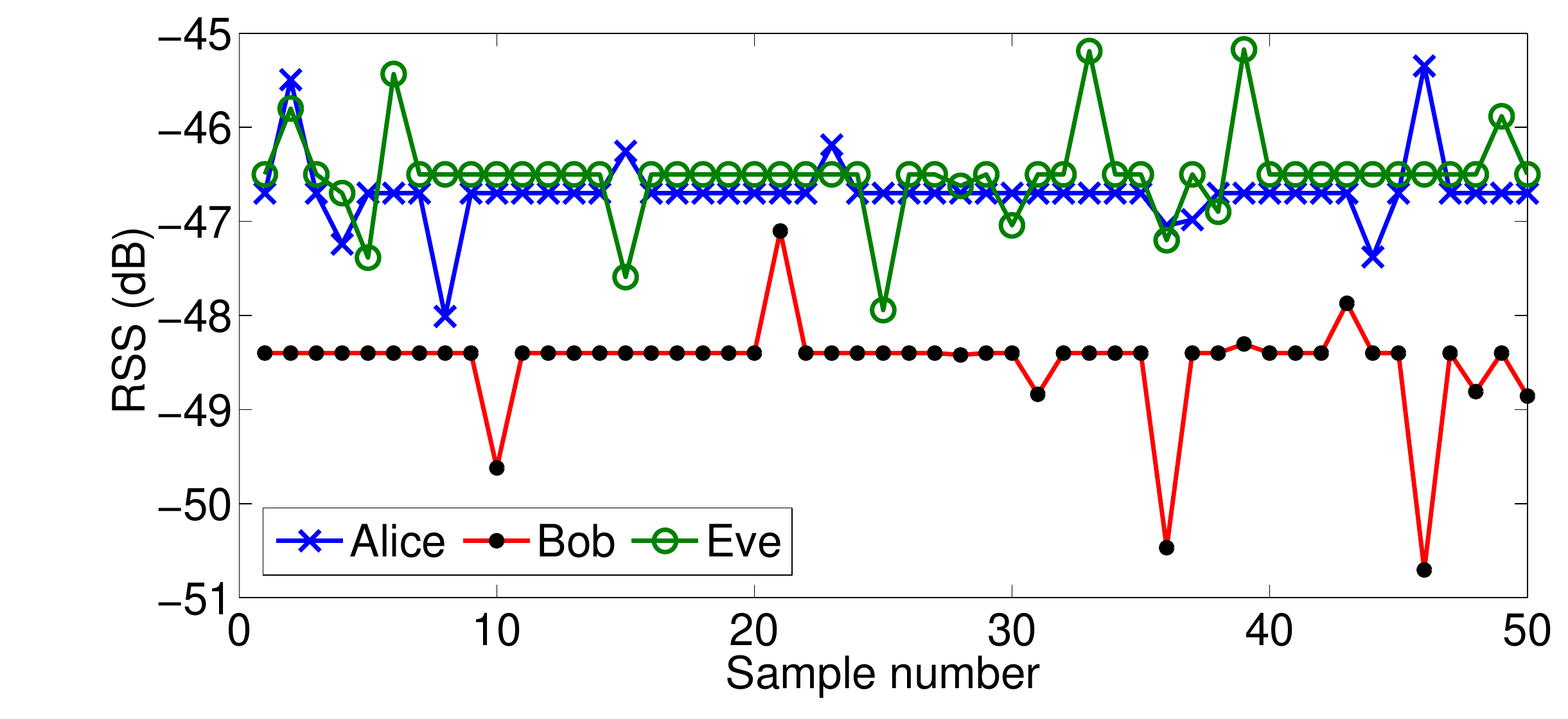}
\end{minipage}}
\caption{Measurements on the playground}
\label{playground}
\end{figure*}

\subsection{Static Endpoints}
We conduct our experiment in static indoor and outdoor scenarios  \emph{B} and \emph{E}
as shown in Figure \ref{office} and Figure \ref{playground}.

Figure \ref{office}(b) and Figure \ref{playground}(b) show the variations of RSS measurements in
an office and on the playground respectively. The channel variation in static
scenarios are less noticeable than those in mobile scenario. We also
note that the curves for Alice and Bob do not follow each other, indicating
a channel with low reciprocity. This happens because the variation
in a static channel is primarily caused by  environment disturbance and
hardware differences which are non-reciprocal. RSS measurements in this
type of environment contain very low inherent entropy.

Figure \ref{office}(a) and Figure \ref{playground}(a) shows the variations of CSI measurements
collected in an office and on the playground respectively.
Though Alice and Bob have different CSI values, their variations trends are similar.
\textbf{The result demonstrates that CSI significantly outperforms RSS on key generation in static scenarios.
On one hand, CSI is more sensitive to environmental changes than RSS.
On the other hand, CSI has a better correlation on channel estimation.}

\begin{figure*}[t]
\begin{minipage}[t]{0.32\linewidth}
\centering
\includegraphics[height=3.8cm]{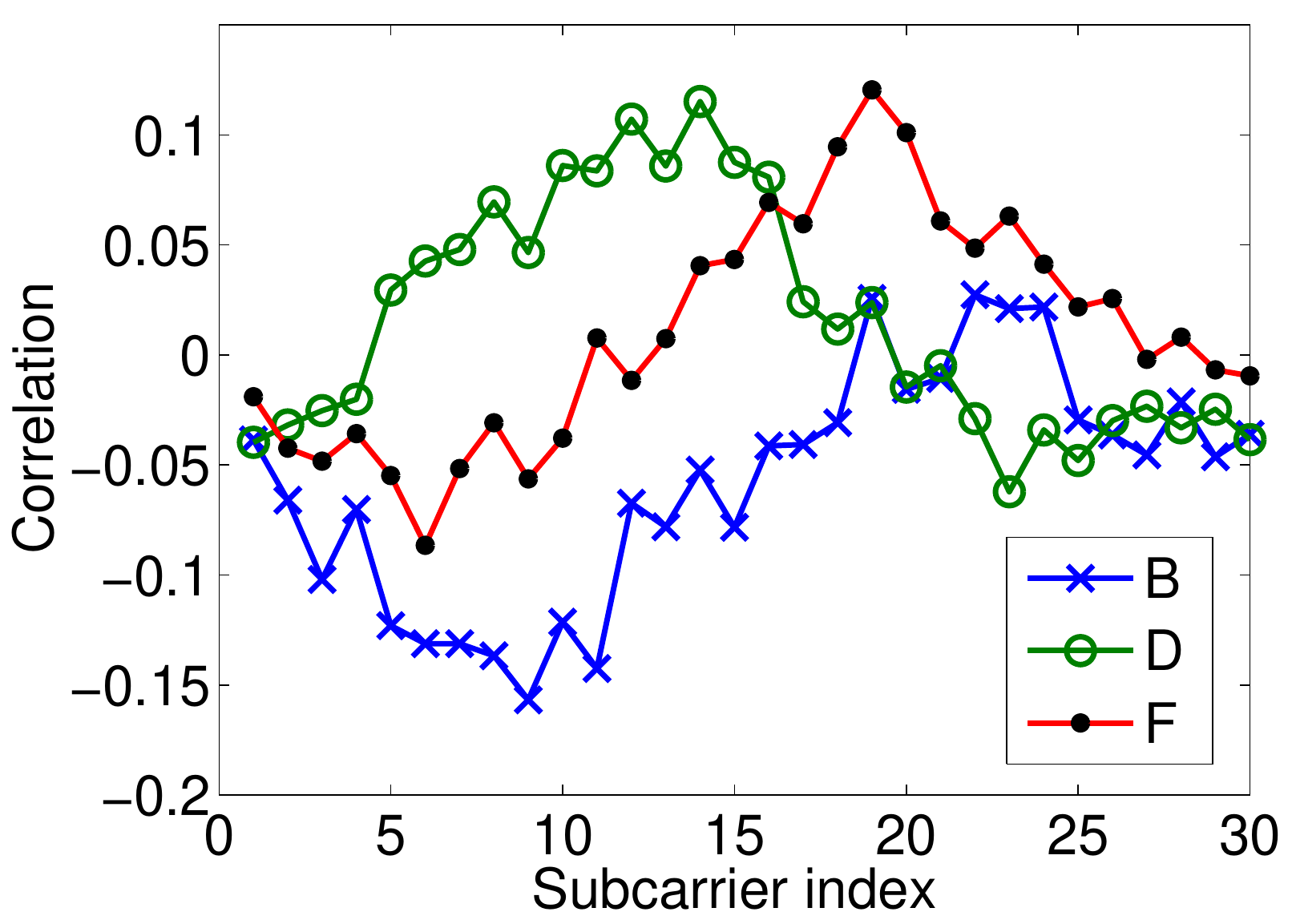}
\caption{The correlation of generated key between Alice and Eve}
\label{evekey}
\end{minipage}
\hfill
\begin{minipage}[t]{0.33\linewidth}
\centering
\includegraphics[height=3.8cm]{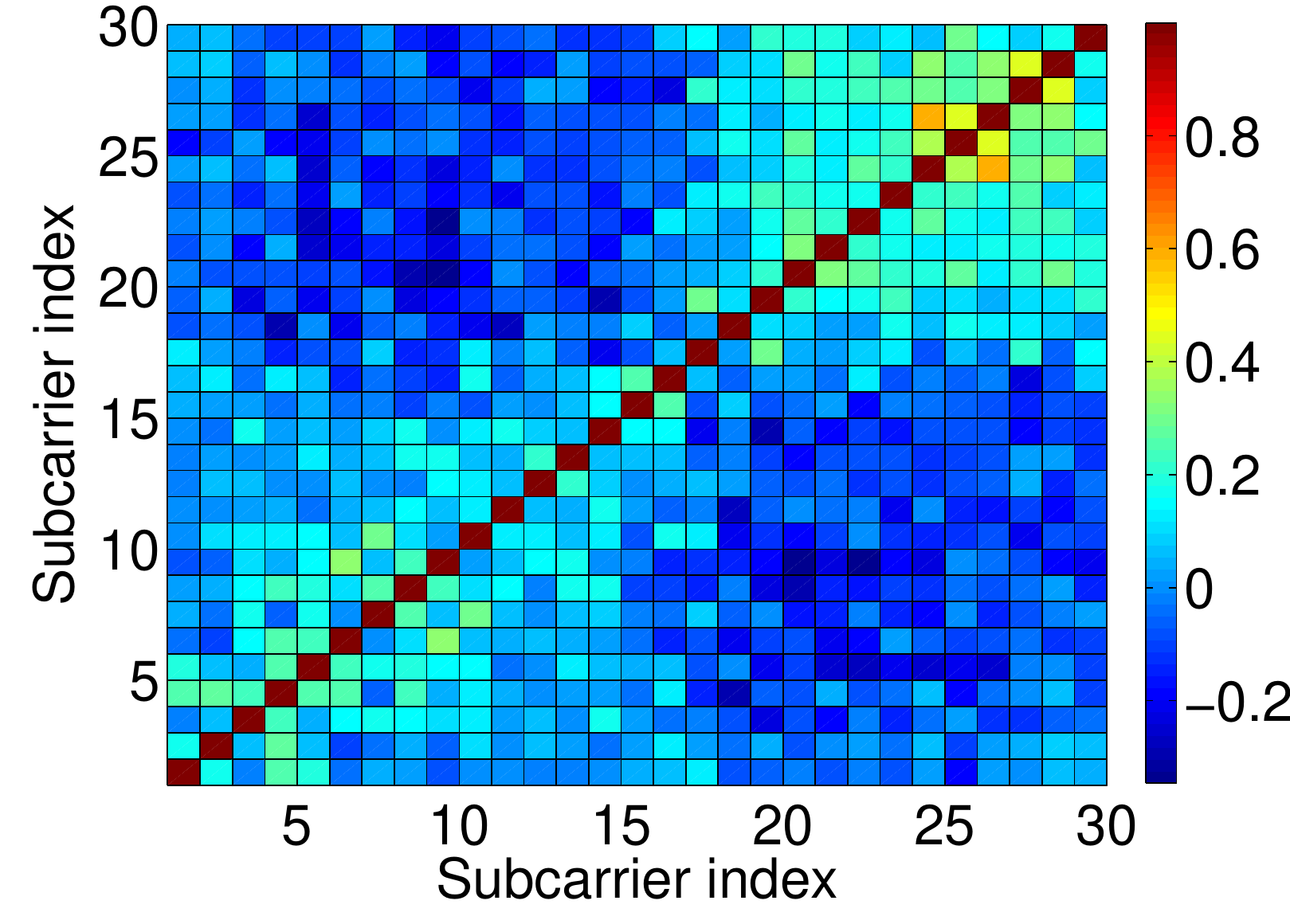}
\caption{The correlation of generated key among 30 subcarriers}
\label{subcarrier}
\end{minipage}
\hfill
\begin{minipage}[t]{0.33\linewidth}
\centering
\includegraphics[height=3.8cm]{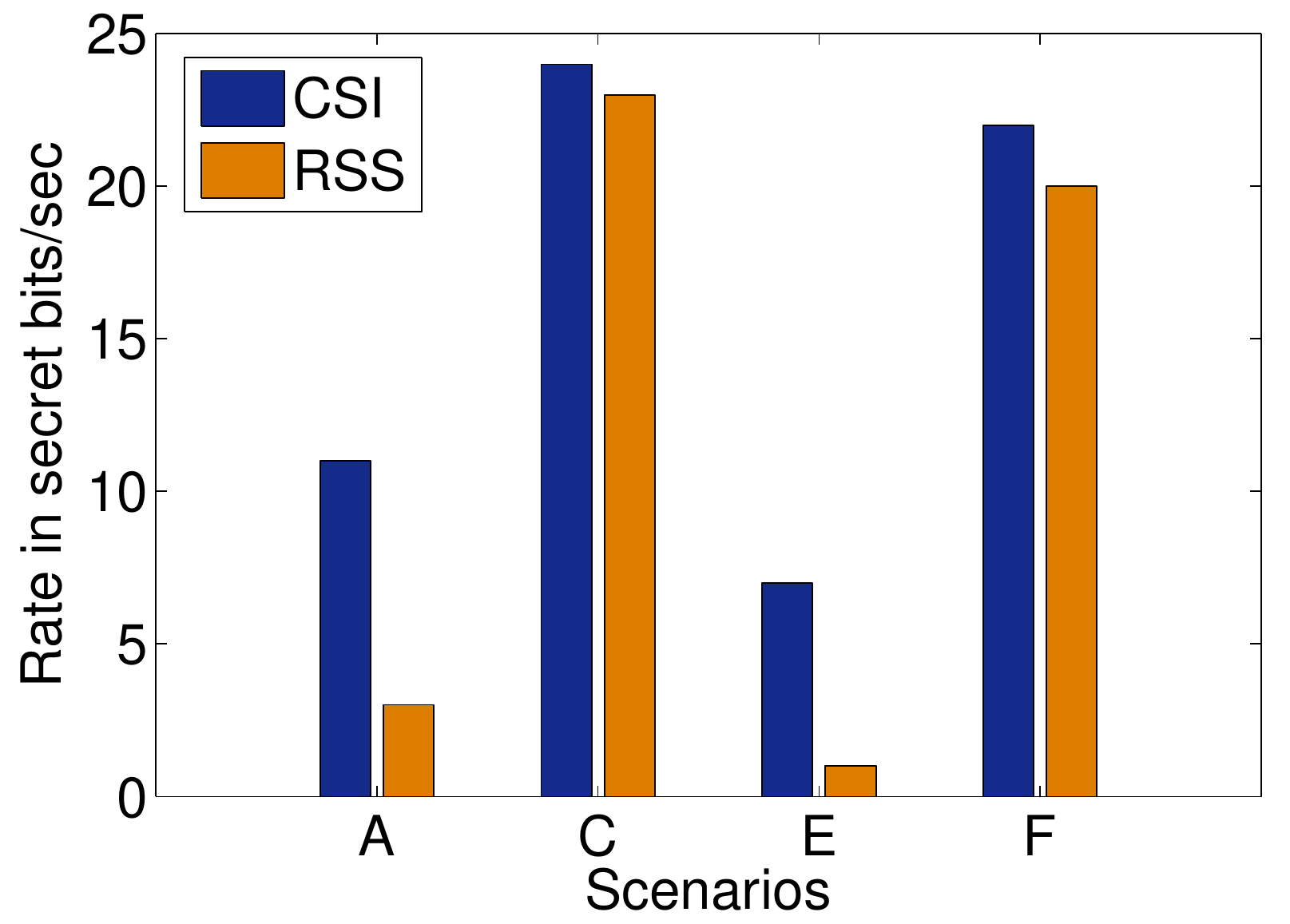}
\caption{Secret bit rate in different scenarios}
\label{rate}
\end{minipage}
\end{figure*}

\begin{figure*}[t]
\subfigure[]{
\begin{minipage}[t]{0.49\linewidth}
\centering
\includegraphics[height=3.8cm]{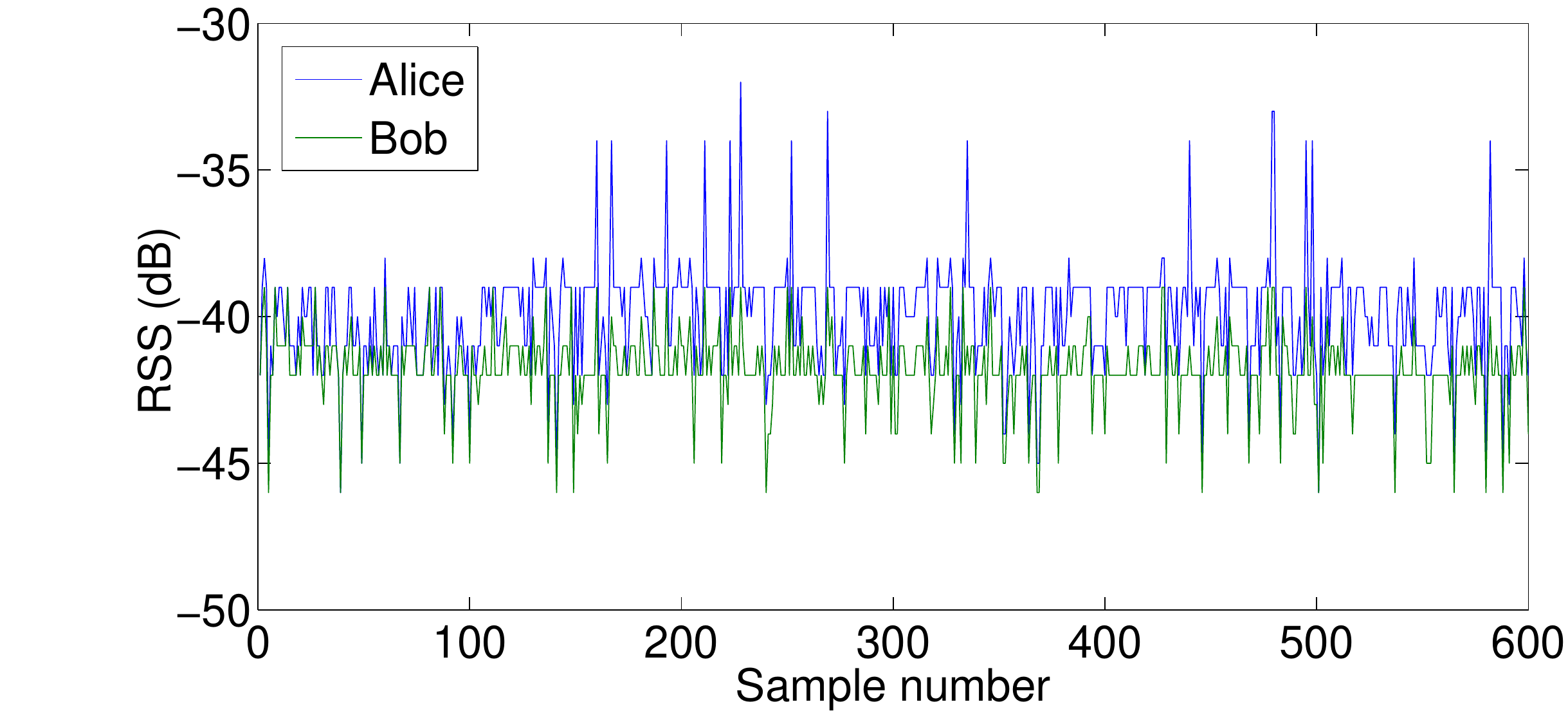}
\end{minipage}}
\hfill
\subfigure[]{
\begin{minipage}[t]{0.49\linewidth}
\centering
\includegraphics[height=3.8cm]{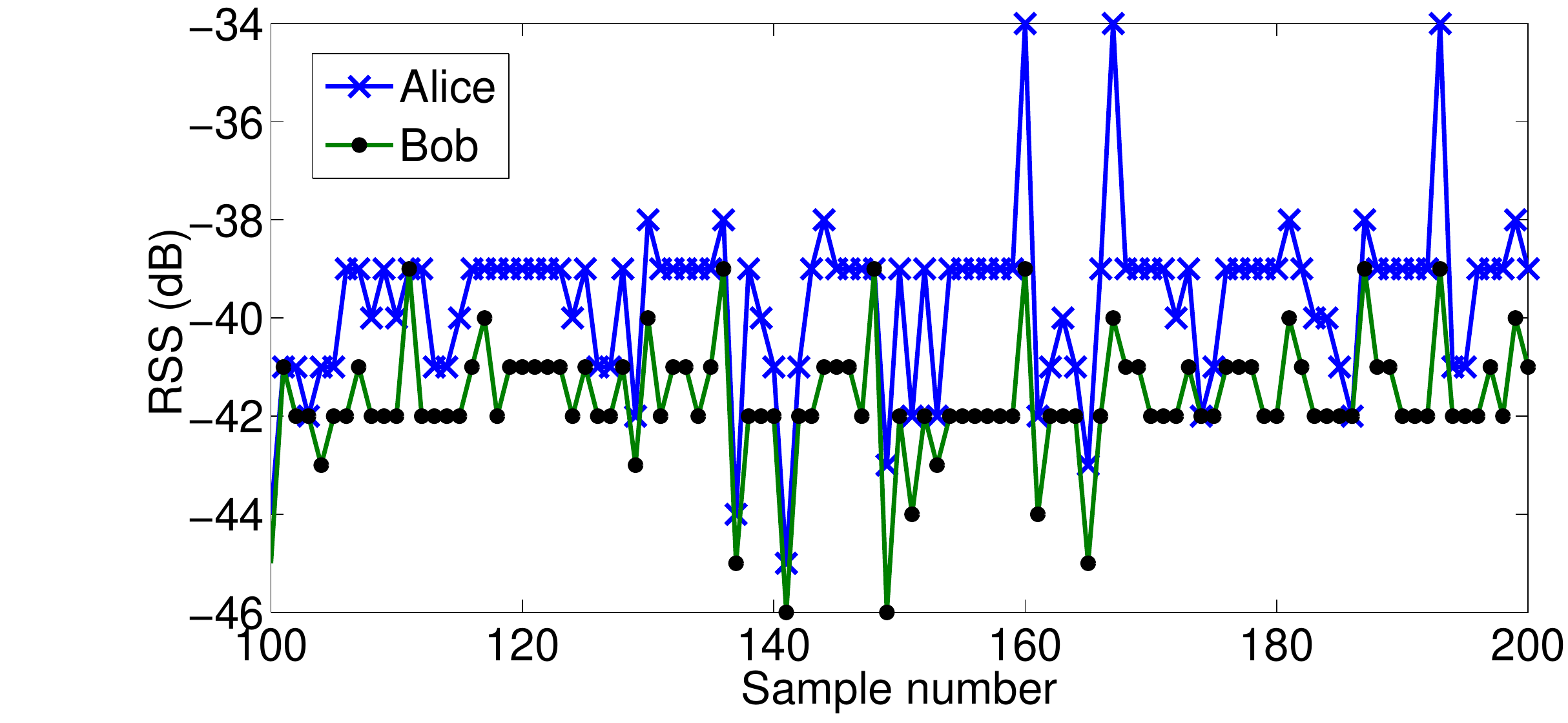}
\end{minipage}}
\caption{(a) Predictable variations of the RSS values, (b) A
magnified portion of the traces.}
\label{pre_rss}
\end{figure*}

\begin{figure*}[t]
\subfigure[]{
\begin{minipage}[t]{0.49\linewidth}
\centering
\includegraphics[height=3.8cm]{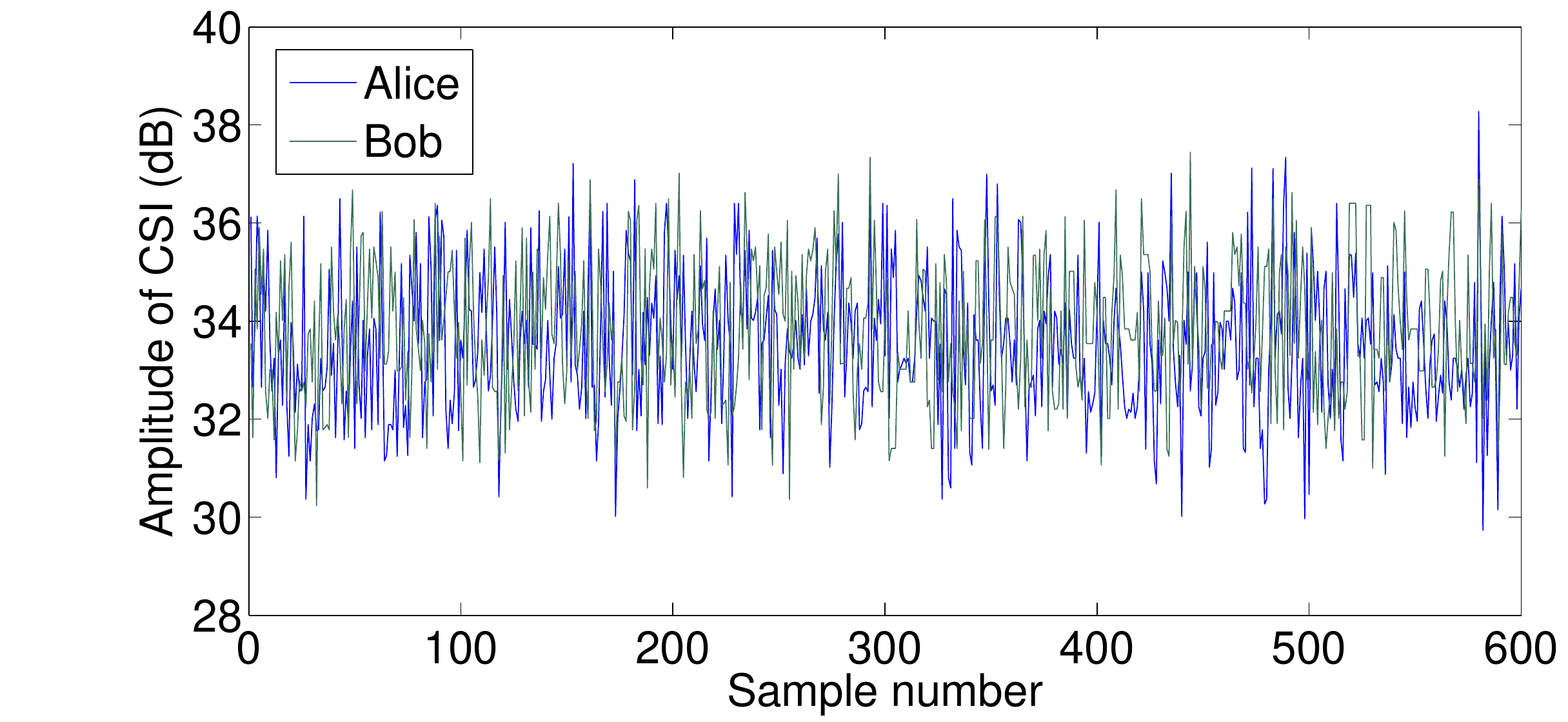}
\end{minipage}}
\hfill
\subfigure[]{
\begin{minipage}[t]{0.49\linewidth}
\centering
\includegraphics[height=3.8cm]{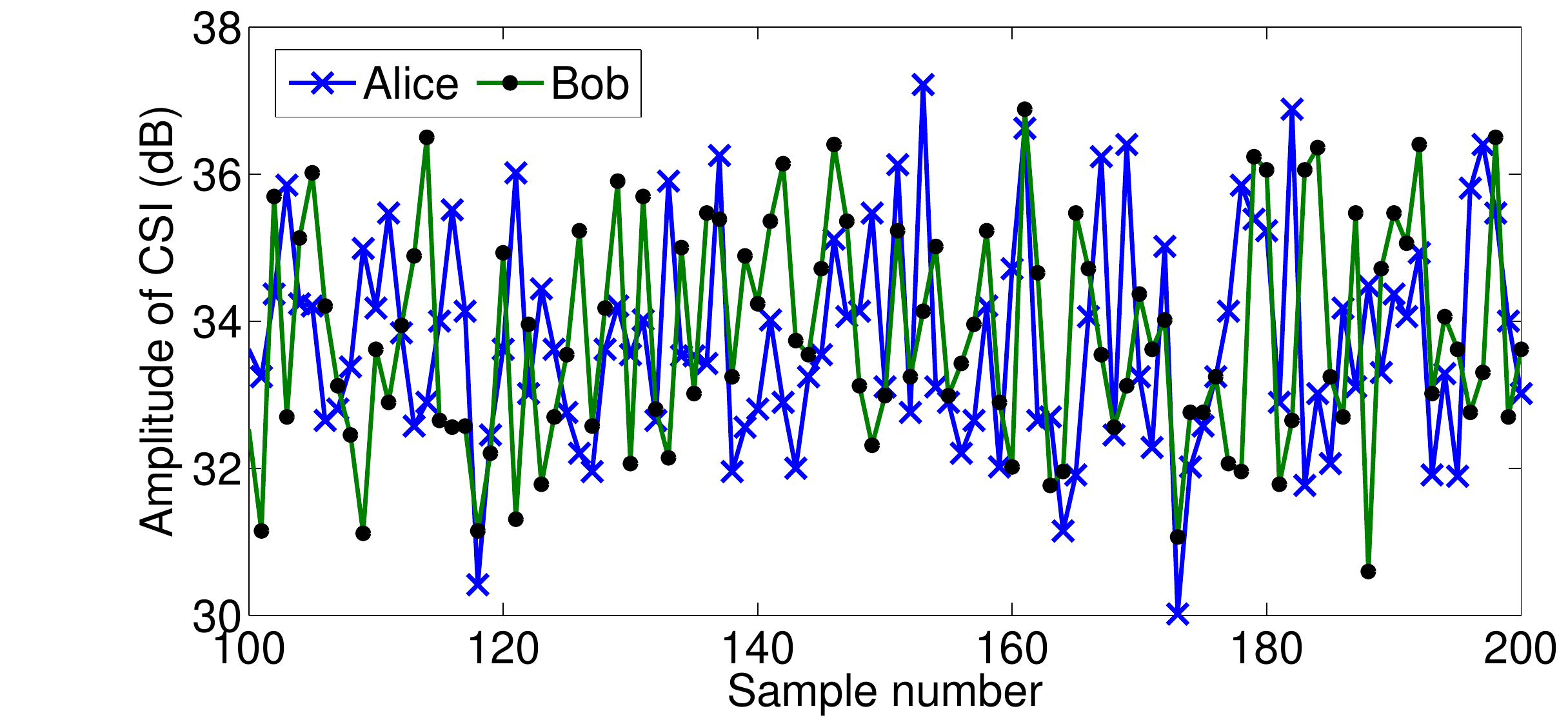}
\end{minipage}}
\caption{(a) The variations of the CSI values, (b) A
magnified portion of the traces.}
\label{pre_csi}
\end{figure*}

\begin{table}\centering
\begin{tabular}{|c|c|c|c|c|c|c|}
\hline
Test                 & A     & B     & C     & D     & E     & F      \\ \hline
Frequency            & 0.73  & 0.57  & 0.84  & 0.38  & 0.12  & 0.75   \\ \hline
Longest run of ones  & 0.32  & 0.11  & 0.16  & 0.88  & 0.67  & 0.21   \\ \hline
FFT                  & 0.60  & 0.73  & 0.95  & 0.34  & 0.79  & 0.68   \\ \hline
Approx. Entropy      & 0.58  & 0.89  & 0.60  & 0.61  & 0.09  & 0.17   \\ \hline
\end{tabular}
\caption{NIST statistical test suite results}
\label{random}
\end{table}
\subsection{Randomness of Key}
Guaranteeing that the generated bits are random is crucial
for key generation.
Since we have assumed the adversary possesses a complete
knowledge of our algorithm, any non-random process in the
bit sequence can be leveraged by the adversary to reduce
the time-complexity of cracking the key. For example, if
there are always more ``1"s than ``0"s in bit stream, then
the effective search space for the adversary would be reduced.
Consequently, a variety of statistical tests have been proposed
to test for various defects.

The $p$-value from each test is listed in Table \ref{random}.
To pass a test, the $p$-value for that test must be greater
than 0.01. We find that the bit streams generated from CSI pass
all the tests.

\subsection{The Correlation of Generated Key between Alice and Eve}
The randomness test results indicate whether SKECE is
secure to defend the key from crackers.
It's crucial for us to make sure whether Eve can generate
the similar key from its channel measurements.

We evaluate the correlation of generated key between Eve and
Alice in different scenarios. The most familiar measure of
dependence between two quantities is ``Pearson's correlation".
It is $+1$ in the case of a perfect positive linear relationship
(correlation), $-1$ in the case of a perfect decreasing (negative)
linear relationship (anticorrelation). As it approaches zero,
there is less of a relationship.

We use Pearson's correlation to estimate the independence
between Eve and Alice in scenarios \emph{B}, \emph{D}, and \emph{E}
as shown in Figure \ref{evekey}. All the values are between $-0.15$
and $+0.15$  for 30 subcarriers in all three scenarios. The
result indicates that Eve's keys are always independent from Alice's key,
even though Eve is 10cm apart from Alice in scenario \emph{D}. It is
because that the received signal rapidly decorrelates over a short
distance. Since he observes a nearly independent channel,
it is impossible for Eve to crack the key from the multipath
fading channel used by Alice.

\subsection{Independence of Bits of 30 Subcarriers}

We extract secret-key based on CSI measurement from 30
subcarriers. The correlation of 30 bit streams of subcarriers
concerns whether all these bit streams can be used as secret-key.

We also evaluate the correlation using Pearson's correlation.
The bit streams are generated in scenario \emph{C}.

Figure \ref{subcarrier} plots the correlations among all 30 subcarriers.
All the correlations are always between $-0.3$ and $+0.3$ except the
correlations with themselves. It states that the key extracted
from 30 subcarriers is independent. OFDM encodes digital data on
multiple carrier frequencies. These sub-carrier signals are
orthogonal with different frequencies. They have the different
channel responses at the receiver in frequency domain. \textbf{The extracted
bit streams from 30 subcarriers are independent with each other, all of
which can be used as secret-keys simultaneously.}

\subsection{Rate of Key Generation}
An important quantity of interest is the rate of generating
secret bits, expressed in secret-bits per second or
¡®s-bits/sec¡¯. Naturally, it is desirable that Alice and Bob
achieve a high secret-bit rate. According to 802.11 recommendations,
it is generally desirable for master keys to be
refreshed at one hour intervals.

Figure \ref{rate} compares the rate of key generation from CSI
and RSS measurements in scenarios \emph{A}, \emph{C}, \emph{E}, and
\emph{F}. It is easy to find that the rate of key generation from CSI
is nearly 4 times higher than RSS in  scenario \emph{A}, and is more than
6 times higher in scenario \emph{C}.

In order to facilitate the comparison, Figure \ref{rate} only shows the
mean of key generation rate of 30 subcarriers. \textbf{Infact, the key generated
from CSI is the sum of all the subcarriers. Thus, the rate of key generation is thirty-fold the
mean value. The rate of key generation from CSI is 32 times higher than
RSS in mobile scenarios, and 100 times higher in static scenarios.}

\subsection{Predictable Channel Attack}
As mentioned earlier, stationary environments reduce the variation of
the channel. An intelligent adversary can use deliberately planned
movements in such scenarios to produce desired and predictable changes
in the channel between the actual sender and receiver.

We conduct an experiment to emulate the predictable channel attack
in the meeting room. Alice and Bob are separated by 3m, they
keep still and probe the channel to generate the bit streams at
10Hz frequency. Eve periodically blocks the line-of-sight transmission
between Alice and Bob. The distance between Alice and Eve is around
1m.

Figure \ref{pre_rss} shows the variation of the RSS values. The RSS
values display periodical changes with the transfer of time.
The RSS drops when Eve block the line of sight path, and then picks
up when Eve move away. The pattern of variation follows the movements.
In this scenario, it produces a predictable pattern of secret bits from
RSS measurements.

Figure \ref{pre_csi} shows the variation of the CSI values. Though its
change also displays periodicity, the variation of bit streams is
still random, which is unpredictable. It is because  CSI is more
sensitive to environment and it will be changed by any variations of
different environmental factors. \textbf{As long as the attacker at a
third location is more than a few wavelengths from either endpoint, it will
not produce a predictable pattern of secret bits from CSI measurements.}

\section{Conclusion}
In this paper, we proposed a protocol, called SKECE, that exploits the
reciprocity of the transfer function of the wireless multipath
channel to establish a common secret key between
two communicating entities. Our protocol obtains a security
advantage from the fact that the channel is random and the
response decorrelates rapidly over distance, which can
defend against a passive eavesdropper as well as an active
adversary attempting predictable channel attack.

We also present the results of a thorough effort to experimentally
validate the feasibility of the wireless channel for
secret key generation. We used off-the-shelf 802.11n cards
for collecting CSI measurements. SKECE
generates secret bits at a high rate from 30 subcarriers both in static and
mobile scenarios. It adopts a weighted key recombination algorithm to
obtain a consistent bit stream without chasing down the mismatched bits,
once all the 30 pairs of bit streams are inconsistent. It reduces the
communication overhead by 50\%.

\bibliographystyle{abbrv}

\begin{thebibliography}{10}

\bibitem{crepaldicsi}
802.11n working group and others. IEEE 802.11n Specification 2009.

\bibitem{amir2001exploring}
Y.~Amir, Y.~Kim, C.~Nita-Rotaru, J.~Schultz, J.~Stanton, and G.~Tsudik.
\newblock Exploring robustness in group key agreement.
\newblock In {\em Proceedings of IEEE ICDCS}, pages 399--408, 2001.

\bibitem{Brassard1994}
G.~Brassard and L.~Salvail.
\newblock Secret-key reconciliation by public discussion.
\newblock In {\em Proceedings of Advances in Cryptology-Eurocrypt}, pages
  410--423. Springer, 1994.

\bibitem{chan2003random}
H.~Chan, A.~Perrig, and D.~Song.
\newblock Random key predistribution schemes for sensor networks.
\newblock In {\em Proceedings of Symposium on Security and Privacy}, pages
  197--213, 2003.

\bibitem{fuzzy-ex}
Y.~Dodis, L.~Reyzin, and A.~Smith.
\newblock Fuzzy extractors: How to generate strong keys from biometrics and
  other noisy data.
\newblock In {\em Proceedings of Advances in Cryptology-Eurocrypt}, pages
  523--540. Springer, 2004.

\bibitem{csitool}
D.~Halperin, W.~Hu, A.~Sheth, and D.~Wetherall.
\newblock Tool release: gathering 802.11 n traces with channel state
  information.
\newblock {\em ACM SIGCOMM Computer Communication Review}, 41(1):53--53, 2011.

\bibitem{hassan1996cryptographic}
A.~Hassan, W.~Stark, J.~Hershey, and S.~Chennakeshu.
\newblock Cryptographic key agreement for mobile radio.
\newblock {\em Digital Signal Processing}, 6(4):207--212, 1996.

\bibitem{heartbeats2011proximate}
T.~Heartbeats.
\newblock Proximate: Proximity-based secure pairing using ambient wireless
  signals.
\newblock {\em IEEE Wireless Communications}, page~8, 2011.

\bibitem{hershey1995unconventional}
J.~Hershey, A.~Hassan, and R.~Yarlagadda.
\newblock Unconventional cryptographic keying variable management.
\newblock {\em IEEE Transactions on Communications}, 43(1):3--6, 1995.

\bibitem{pr-one-way}
R.~Impagliazzo, L.~Levin, and M.~Luby.
\newblock Pseudo-random generation from one-way functions.
\newblock In {\em Proceedings of ACM STOC}, pages 12--24, 1989.

\bibitem{effkeyex}
S.~Jana, S.~Premnath, M.~Clark, S.~Kasera, N.~Patwari, and S.~Krishnamurthy.
\newblock On the effectiveness of secret key extraction from wireless signal
  strength in real environments.
\newblock In {\em Proceedings of ACM MobiCom}, pages 321--332, 2009.

\bibitem{lee2006distributed}
P.~Lee, J.~Lui, and D.~Yau.
\newblock Distributed collaborative key agreement and authentication protocols
  for dynamic peer groups.
\newblock {\em IEEE/ACM Transactions on Networking}, 14(2):263--276, 2006.

\bibitem{li2010data}
M.~Li, W.~Lou, and K.~Ren.
\newblock Data security and privacy in wireless body area networks.
\newblock {\em IEEE Wireless Communications}, 17(1):51--58, 2010.

\bibitem{liu2005establishing}
D.~Liu, P.~Ning, and R.~Li.
\newblock Establishing pairwise keys in distributed sensor networks.
\newblock {\em ACM Transactions on Information and System Security},
  8(1):41--77, 2005.

\bibitem{ecd}
Y.~Liu, S.~Draper, and A.~Sayeed.
\newblock Exploiting channel diversity in secret key generation from multipath
  fading randomness.
\newblock {\em IEEE Transactions on Information Forensics and Security},
  PP(99):1, 2012.

\bibitem{radio-telepathy}
S.~Mathur, W.~Trappe, N.~Mandayam, C.~Ye, and A.~Reznik.
\newblock Radio-telepathy: extracting a secret key from an unauthenticated
  wireless channel.
\newblock In {\em Proceedings of ACM MobiCom}, pages 128--139, 2008.

\bibitem{Maurer2003}
U.~Maurer and S.~Wolf.
\newblock Secret-key agreement over unauthenticated public channels - part iii:
  Privacy amplification.
\newblock {\em IEEE Transactions on Information Theory}, 49(4):839--851, Apr.
  2003.

\bibitem{wallace2010automatic}
J.~Wallace and R.~Sharma.
\newblock Automatic secret keys from reciprocal mimo wireless channels:
  Measurement and analysis.
\newblock {\em IEEE Transactions on Information Forensics and Security},
  5(3):381--392, 2010.

\bibitem{wilhelm2009key}
M.~Wilhelm, I.~Martinovic, and J.~Schmitt.
\newblock On key agreement in wireless sensor networks based on radio
  transmission properties.
\newblock In {\em Proceedings of IEEE Workshop on Secure Network Protocols},
  pages 37--42, 2009.

\bibitem{xiao2008using}
L.~Xiao, L.~Greenstein, N.~Mandayam, and W.~Trappe.
\newblock Using the physical layer for wireless authentication in time-variant
  channels.
\newblock {\em IEEE Transactions on Wireless Communications}, 7(7):2571--2579,
  2008.

\bibitem{itsk}
C.~Ye, S.~Mathur, A.~Reznik, Y.~Shah, W.~Trappe, and N.~Mandayam.
\newblock Information-theoretically secret key generation for fading wireless
  channels.
\newblock {\em IEEE Transactions on Information Forensics and Security},
  5(2):240 --254, june 2010.

\bibitem{zhou1999securing}
L.~Zhou and Z.~Haas.
\newblock Securing ad hoc networks.
\newblock {\em IEEE Network}, 13(6):24--30, 1999.

\bibitem{zhu2004interleaved}
S.~Zhu, S.~Setia, S.~Jajodia, and P.~Ning.
\newblock An interleaved hop-by-hop authentication scheme for filtering of
  injected false data in sensor networks.
\newblock In {\em Proceedings of IEEE Symposium on Security and Privacy}, pages
  259--271, 2004.

\bibitem{zhu2003establishing}
S.~Zhu, S.~Xu, S.~Setia, and S.~Jajodia.
\newblock Establishing pairwise keys for secure communication in ad hoc
  networks: A probabilistic approach.
\newblock In {\em Proceedings of IEEE ICNP}, pages 326--335, 2003.

\end{thebibliography}

\end{document}